\documentclass{ws-ijmpa}
\newcommand\simlt{\lower.5ex\hbox{$\; \buildrel < \over \sim \;$}}
\newcommand\simgt{\lower.5ex\hbox{$\; \buildrel > \over \sim \;$}}
\begin{document}

\markboth{Amir Levinson}
{High-Energy Aspects of Astrophysical Jets}

%
\catchline{}{}{}{}{}
%

\title{HIGH-ENERGY ASPECTS OF ASTROPHYSICAL JETS}

\author{AMIR LEVINSON}

\address{School of Physics and Astronomy, Tel Aviv University\\
Tel Aviv 69978, Israel}

\maketitle

\begin{history}
\received{8 November 2006}
\end{history}

\begin{abstract}
Various aspects of the high-energy emission from relativistic jets associated with compact astrophysical systems are reviewed.  
The main leptonic and hadronic processes responsible for the production of high-energy $\gamma$-rays,
very-high energy neutrinos and ultra-high energy cosmic rays are discussed.   Relations between the $\gamma\gamma$ pair production and photomeson production opacities are derived, and their consequences for the relative emission of $\gamma$-rays and neutrinos are examined. 
The scaling of the size and location of the various emission zones and other quantities with black hole mass and dimensionless luminosity is elucidated.   The results are applied to individual classes of objects, including blazars, microquasars and gamma-ray bursts.  It is concluded that if baryons are present in the jet at sufficient quantities, then under optimal conditions most systems exhibiting relativistic jets 
may be detectable by upcoming neutrino telescopes.  An exception is the class of TeV blazars, for which $\gamma$-ray observations imply neutrino yields well below detection limit.  

\end{abstract}


\section{Introduction}	
It is widely believed that the high-energy emission observed in 
several classes of Galactic and extragalactic sources, e.g., blazars, 
microquasars, gamma-ray bursts (GRBs), is associated with collimated, 
relativistic outflows.  This view is strongly supported by some recent 
exciting discoveries, most notably (i) the discovery that many of the 
compact extragalactic radio sources are  extremely luminous, hard $\gamma$-ray 
sources,  (ii) detection of superluminal motions and very-high energy $\gamma$-ray emission 
in several Galactic X-ray transients, (iii) indirect evidence for jets in GRBs,
and (iv) detection of a mildly relativistic outflow from a magnetar following a giant flare (GF).
An overview of the observations is given in \S \ref{sec:observations} below.

A common view is that relativistic jets associated with compact astrophysical 
systems are powered by a magnetized accretion disk and a spinning 
black hole and collimated by magnetic fields and/or the medium surrounding the jet.  
The outflows associated with highly super-Eddington sources are presumably driven by 
violent events, specifically collapse of a massive star or coalescence
of compact objects in the case of GRBs, and magnetic field annihilation in the case of magnetars.  
Unfortunately, the mechanism responsible for the formation, acceleration and collimation
of the jets is poorly understood at present.  Even the composition of the jet is unknown in most sources.
The high-energy emission exhibited by those systems is believed to be produced 
close to the inner engine, in regions where dissipation of the bulk (kinetic plus magnetic) 
outflow energy occurs and, therefore, provides an important probe of the inner jet zone.  As discussed below, 
some constraints on jet parameters can be imposed from the observations.
Further progress in our understanding of the physics of relativistic jets and their engines is expected
in the coming few years with the advance of observational techniques.
A new generation of experiments just started operating or will become operative soon:
Space-based (e.g. GLAST, INTEGRAL, AGILE), Cerenkov
(e.g. HESS, MAGIC, VERITAS), and air-shower (e.g. MILAGRO)
$\gamma$-ray detectors will probe with high sensitivity the energy
range of 10~MeV to a few TeV (see, e.g., Ref.~\refcite{Catanese99} for reviews). 
The operating BAIKAL and AMANDA neutrino telescopes,
and the cubic-km scale telescopes under construction (IceCube,
ANTARES, NESTOR, NEMO; see, e.g. Ref.~\refcite{Halzen05} for review),
will open a new window onto the Universe. Besides providing an important probe of the 
innermost regions of compact astrophysical systems, these experiments can also be exploited to 
test new physics.   Finally, new ultra-high-energy cosmic-ray detectors, e.g. the 
HiRes \cite{HiRes} and the hybrid Auger detectors will provide cosmic-ray data of unprecedented quality and quantity. 

\subsection{The Basic Model}	
\begin{figure}[pb]
\centerline{\psfig{file=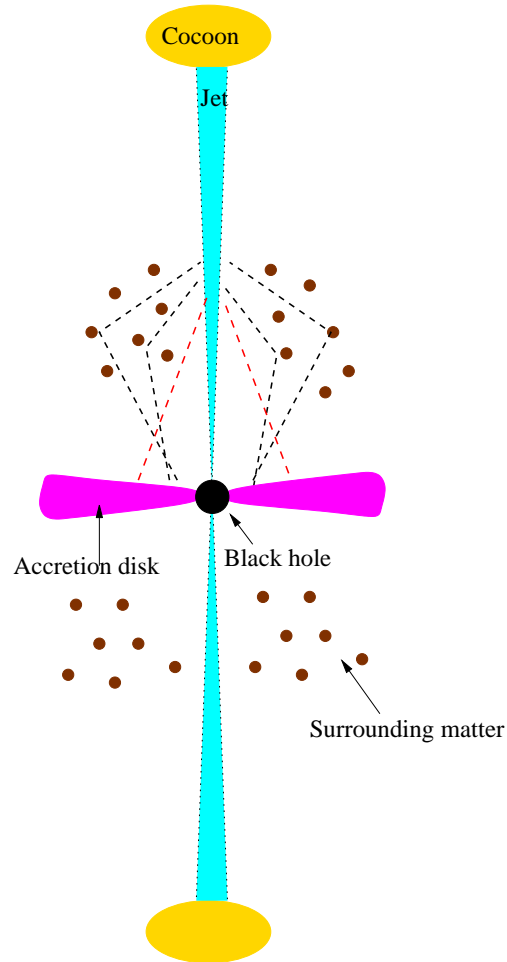,width=6.7cm}}
\vspace*{8pt}
\caption{Schematic illustration of the high-energy emitting jet model.  A relativistic jet is ejected by a central engine consisting of a black hole surrounded by a hot accretion disk.  Soft X-ray photons emitted from the inner disk radii (indicated by the dashed lines) may be intercepted directly by the jet (red lines) and/or Thomson-scattered across the jet by matter surrounding it (black lines).  External photons penetrating the jet interact with relativistic electrons and protons accelerated in situ, resulting in emission of high-energy $\gamma$-rays and very high-energy neutrinos.  Synchrotron emission by the relativistic electrons accelerated in the jet contributes additional source of seed photons.  As the jet propagates through the ambient (interstellar of intergalactic) medium it drives a strong shock that decelerates it.  The shocked ambient gas forms a hot cocoon that gives rise to additional emission from much larger scales.  The radio lobes seen in FR2 sources and microquasars, and the afterglow emission seen in long GRBs are clear diagnostics of the shocked ambient gas.}
\label{fig:disk-ext}
\end{figure}

\begin{figure}[pb]
\centerline{\psfig{file=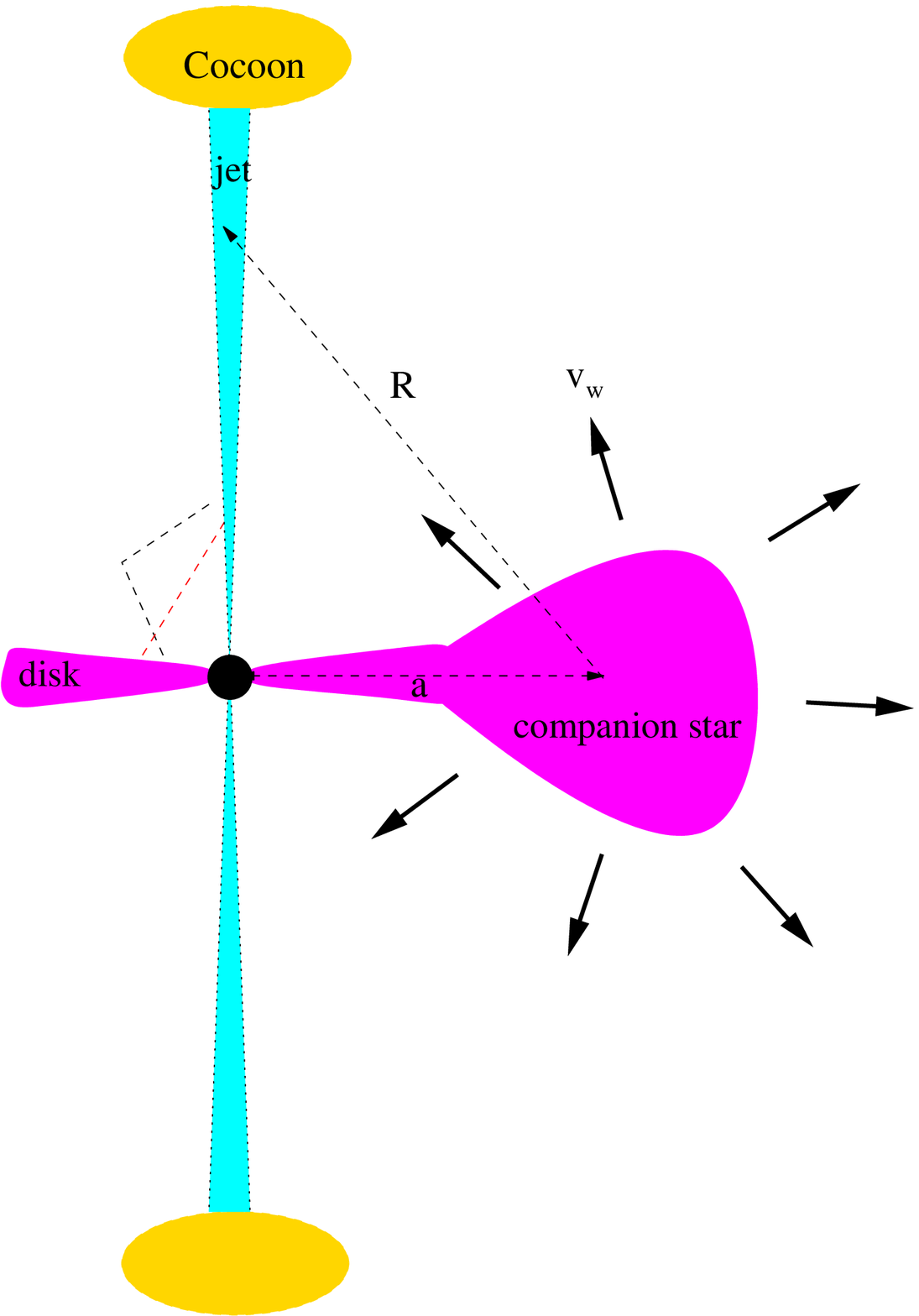,width=7.7cm}}
\vspace*{8pt}
\caption{A sketch of a microquasar system.}
\label{fig:MQSW}
\end{figure}

The system under consideration is depicted in Fig. \ref{fig:disk-ext}.  The main ingredients
relevant to the discussion that follows are indicated.   In blazars and microquasars radiation emerging 
from the disk (represented by the dashed lines in Fig. \ref{fig:disk-ext}) may be an important source of seed photons in the jet.
In GRBs the disk is highly opaque to electromagnetic radiation and cools through emission of MeV neutrinos.  Moreover, 
the jet must first break out of the envelope of the progenitor star, driving a strong shock.  The break-out shock  
may give rise to additional emission of $\gamma$-rays and neutrinos at early times, as will be described below.  In microquasars a companion star may be present, as illustrated in Fig. \ref{fig:MQSW}.  Its radiation and stellar wind can interact with the jet and may contribute additional opacities.

\section{Overview of Observations}
\label{sec:observations}
\subsection{Blazars}
\label{sec:obs-blazars}

Many compact extragalactic radio sources, called blazars, have been found by 
the {\it Compton} Gamma Ray Observatory (CGRO) to be extremely luminous, hard $\gamma$-ray 
sources\cite{Fichtel94}\cdash\cite{Hartman99}.  The compact radio sources 
are members of a larger class of sources (constituting $\sim 10$ \% of all AGNs), 
designated as radio loud, that exhibit radio jets extending over many decades in radius
(see Ref.~\refcite{Begelman84} for a review). 
The rapid variability and the superluminal motion of radio knots often seen in blazars
indicate that those jets expand relativistically.  According to the unified model \cite{Urry/Padovani95}, 
both compact and extended radio sources belong to the same class of physical objects, distinguished
observationally by orientation to the observer; jets pointing in our direction are 
classified as blazars.   

Nearly a hundred blazars (66 high-confidence identifications and 27 
potential identifications) are listed in the 3rd EGRET catalog\cite{Hartman99} as
0.1 -  20 GeV $\gamma$-ray sources.   Their distances and $\gamma$-ray luminosities span a wide
range, with the most powerful sources (e.g., 0528+134, 4C38.41) exhibiting isotropic equivalent $\gamma$-ray luminosities 
as high as $10^{49}$ erg s$^{-1}$.  Despite observational
efforts no extended radio sources or radio-quiet AGNs have been detected at $\gamma$-ray
energies (except for a marginal detection of Cen A).  This apparent exclusive association of EGRET AGN sources with compact radio sources strongly supports the unified model, suggesting that the 
$\gamma$-rays are produced inside the jet and are beamed, with a beaming factor
$f_b\simeq\theta^2/2\sim10^{-3}-10^{-2}$, where $\theta$ is the jet semi-opening angle.                        
The energy spectra in the range 0.1-10 GeV can be                                                                   
fitted by power laws, with energy spectral indices $\alpha_\gamma$ in                                            
the range 0.7 to 1.4, (though the broad energy response of the instrument
coupled with the paucity of photons cannot really exclude spectral 
curvature).  

Some blazars also exhibit very high energy (VHE) $\gamma$-ray emission.  To date about a dozen 
blazars have been reported as VHE ($>100$ GeV) sources\cite{Punch92}.
The energy spectra above 100 GeV are best fitted by power laws with upper cutoff at around a few TeV.
Essentially all of the VHE sources are high-frequency-peaked BL Lac (HBL) objects
located at relatively low redshifts ($z<0.2$).   
The only exception is the FRI radio galaxy M87\footnote{Although M87 is believed to be a miss-aligned 
BL Lac the origin of the VHE $\gamma$-rays is unclear, as the emission from the jet 
is expected to be beamed away from us}. The BL Lac objects are blazars that show very weak emission
lines or no emission lines at all\footnote{Technically, a blazar is defined as BL Lac if the line equivalent
width is less than 5 Angstroms}, in contrast to the flat-spectrum radio quasars (FSRQ)
that exhibit broad optical emission lines.  This suggests that in BL Lac objects disk emission
is strongly suppressed.  

The broad band $\nu F_\nu$ spectrum of blazars is characterized by two main 
spectral components.  A low energy component 
peaking in the submm-to-IR regime in FSRQs and in the UV-to-hard X-ray band in 
HBLs, and a high energy one peaking at MeV energies in FSRQs and GeV energies in HBLs.   
The ratio of luminosities of the high-and low-energy spectral components is
typically large in the powerful sources, and is of order unity in the 
BL Lac objects (but may change during outbursts).  
The beamed radio-to-UV emission is most likely synchrotron radiation by non-thermal 
electrons (the synchrotron spectrum may extend up to hard X-ray 
energies in BL Lac objects).  The origin of
the high-energy (hard X-ray/$\gamma$-ray) emission is still debatable.  It may be due 
to inverse Compton scattering of  
some seed radiation by electrons accelerated in situ, or it may have hadronic 
origin\cite{mannheim93}.  Conceivable  
sources of scattered photons include the (beamed) synchrotron  
radiation produced in the jet (SSC mechanism; e.g., Refs.~\refcite{Konigl81}
--\refcite{Konopelko03}) or external radiation (ERC mechanism; e.g, Refs.
~\refcite{Begelman87}--\refcite{Blandford/Levinson95}),
 presumably continuum radiation emanating from the inner parts
of an accretion disk, that is either directly enters the jet 
or, alternatively, reprocessed and/or scattered across the
jet by gas surrounding it, as illustrated in Fig.~\ref{fig:disk-ext}.

Strong, rapid variability over the entire
electromagnetic spectrum is one of the characteristics of blazar emission. 
Large amplitude variations of the 0.1-10 GeV $\gamma$-ray flux 
on timescales of days or weeks are typical to many EGRET 
blazars\cite{Kniffen93,Fichtel/Thompson94}.
Doubling times as short as a few hours have been
reported for some of the strongest flares\cite{Mattox97,Wehrle97}. 
It should be noted that the variability time that could 
be measured at GeV energies at the time was limited by the sensitivity of 
EGRET, so it is conceivable that the high-energy emission 
in blazars varies on even shorter timescales.  Indeed, faster flux variations
has been observed in the TeV blazars.  In the case of Mrk 421, 
flux doubling time as short as 15 min has been reported\cite{Buckley96}.  
This rapid variability of the hard $\gamma$-ray 
emission provides severe  constraints on the bulk Lorentz factor of the 
outflow, and on the compactness and location of the emission region. 
It also has important consequences for the production of VHE neutrinos
that will be discussed in \S \ref{sec:application}.

Radio observations of EGRET AGN sources have shown that in 
many cases radio outbursts seem to follow $\gamma$-ray flares with time lags of weeks
to months\cite{Reich93,Zhang94}.  In a one or two exceptional 
cases, however, sources which 
have been monitored at radio frequencies for long time showed no 
radio activity following an EGRET flare.
In general, time lags between $\gamma$-ray and radio outbursts are 
expected if the $\gamma$-ray emission region is located well within the 
radio photosphere (see Fig. \ref{fig-scales}) .  Nonetheless, more complicated 
relationship can be envisioned.  For example, in the radiative front model 
the delay between radio and $\gamma$-ray outbursts can be controlled by
changing the opacity at the creation radius.  In that case, one may 
expect to have some correlations between the ratio of amplitudes of 
the variations and the delay time\cite{Eldar00}.   

The relation between the $\gamma$-ray and optical/UV fluxes is not 
well characterized yet.  In some sources the optical, UV, X-and $\gamma$-ray fluxes 
seem to be correlated given the temporal resolution\cite{Wagner96,Wehrle97}.
In the quasar 3C279 there are some indications for lags of 1 to 2 
days between the UV and $\gamma$-ray peak\cite{Wehrle97}, 
but these are not statistically significant.  Future multi-wavelength observations may be able to resolve
such short time lags, and provide additional constraints on the models.
Observations of the BL Lac Mrk 421 reveal simultaneous X-ray/TeV 
outbursts with no significant variations of the flux at EGRET 
energies\cite{Macomb95,Takahashi96}.  Such an event has no natural explanation
within the framework of inhomogeneous ERC models.  It may be reproduced by the one-zone SSC model
under certain assumptions\cite{Mastichiadis/Kirk97}.

\subsection{Microquasars and $\gamma$-Ray Binaries}

Microquasars are Galactic X-ray binary (XRB) systems, which exhibit 
relativistic radio jets \cite{Mirabel94}\cdash\cite{Fender01b}.
These systems are believed to consist of 
a compact object, a neutron star or a black hole, and a giant  
star companion. Mass transfer from the giant star to the compact 
object through the formation of an accretion disk and the presence 
of the jets make them similar to small quasars, hence their name  
``microquasars.'' The analogy may not be only morphological; although
there is no obvious scaling, it is common thinking that the physical processes 
that govern the formation of the accretion disk and the ejection of plasma into the 
jets are the same for both systems. Local, Galactic 
microquasars may therefore be considered as nearby ``laboratories,'' 
where models of the distant, powerful quasars can be tested.  

There are, nonetheless, some important environmental differences that can affect the 
resulting high-energy emission from the system.  In particular, in microquasars
with a high-mass stellar companion (HMXBs) the radiation and the stellar wind  
emanating from the companion star can interact with jet, leading to 
additional emission of $\gamma$-rays and neutrinos.  The compact object in some of those HMXB systems 
may be a pulsar.  In this case high-energy emission may result from the interaction of 
the pulsar wind with the stellar radiation.  Those systems are termed now $\gamma$-ray binaries\cite{Dubus}.
The emission from HMXBs is expected to be modulated, owing
to the rotation of the binary system.

The temporal behavior of microquasars appears to be rather complex.  They
exhibit large amplitude variations over a broad range of time scales
and frequencies, with apparent connections between the radio, IR, and 
soft/hard X-ray fluxes \cite{Harmon97}\cdash\cite{Yadav01}.  The characteristics of 
the multi-waveband behavior depend on the state of the source, that is, whether 
the source is in a very high, soft/high or low/hard state \cite{Fender01}.  The 
ejection episodes are classified into several classes according to the brightness 
of synchrotron emission produced in the jet and the characteristic time scale 
of the event \cite{Eikenberry00}.  The duration of 
major ejection events (class A) is typically on the order of days, while
that of less powerful flares (classes B and C) is correspondingly shorter
(minutes to hours).  The correlations between the X-ray and synchrotron emission clearly indicates 
a connection between the accretion process and the jet activity.  Whether 
radio and IR outbursts represent actual ejection of blobs of plasma 
or, alternatively, formation of internal shocks in a quasi-steady jet is 
unclear at present (cf. Ref.~\refcite{KSS00}).

$\gamma$-ray emission from microquasars has been predicted shortly after their 
discovery \cite{Levinson/Blandford96,Atoyan/Aharonian99}.
Early attempts to detect microquasars with EGRET yielded only 
upper limits \cite{Levinson/Mattox96}. Tentative identifications of two EGRET sources 
with the  HMXBs LSI+61 303\cite{Kniffen97} and  LS 5009 \cite{Paredes00} 
were reported later.  Both systems have been detected recently by TeV observatories,
confirming the EGRET identifications\cite{Aharonian05a}.  Recent observations suggest that 
LSI +61 303 is a $\gamma$-ray binary.
Possible association of some unidentified, variable
EGRET sources with Galactic black hole systems has been noted in Ref.~\refcite{Torres01}.

\subsection{Gamma Ray Bursts}
\label{sec:obs-GRB}
The observed isotropic equivalent energy $E_{iso}$ of GRBs span the range $10^{50}-10^{54}$ ergs
(see e.g., Refs.~\refcite{Piran05} for recent reviews on GRBs).  A large fraction of this energy is released
during the prompt emission phase that, in long duration GRBs lasts for about several tens of seconds.
This phase is followed by a longer phase during which {\it afterglow} emission at X-ray-to-radio
wavelengths is typically observed.
The spectrum of the prompt emission can be fitted by a broken power law\cite{Band93}, with a peak energy
that appears to be correlated with $E_{iso}$\cite{Amati02} (referred to as Amati relation), albeit with a large 
scatter (but c.f., Ref.~\refcite{Nakar03}), and is about 1 MeV in the most luminous sources.  In many
sources the spectrum extends well above the peak, up to an energy of $\sim100$ MeV in some cases, 
indicating that the emission cannot be produced in an adiabatically expanding e$^\pm$ fireball, 
as originally proposed\cite{Paczynsky86}.  Compactness arguments further
imply that in sources exhibiting emission up to $\sim100$ MeV the Lorentz factor must 
exceed 100 or so.  The large values of $E_{iso}$ observed and the indications for achromatic breaks in the 
lightcurves of afterglow emission\cite{Kulkarni99} are suggestive evidence for collimation of GRB 
emitting outflows\cite{Kulkarni99,Roads97}.  The distribution of opening angles inferred from the 
break of the afterglow lightcurves peaks at 
$\theta\sim0.1$\cite{Frail01}.  Beaming corrections seem to significantly reduce the scatter in the 
Amati relation\cite{Ghirlanda05}, suggesting that this scatter is predominantly
due to the spread in opening angles of the GRB population.  It has been proposed that the correlation is
entirely due to viewing angle effects and that the {\it true} energy of long GRBs may be 
standard with $E_j\sim10^{51}$ ergs\cite{Eichler04}. 

The composition of GRB jets is yet an open issue.  The standard view until recently was that
the explosion energy is converted to kinetic energy of baryonic shells that collide and 
drive internal shocks at radii $r\sim 10^{12}-10^{13}$ cm, behind which the prompt emission
is produced.  Production of UHECRs\cite{Levinson/Eichler93,Waxman95} and VHE 
neutrinos\cite{Waxman/Bahcall97} is then expected under certain conditions.  
Alternative scenarios have been proposed (e.g., Ref.~\refcite{Ghisellini99}),
in which the MeV peak is limited by pair production.  The recent detections of an early, shallow 
afterglow phase and sharply rising, large amplitude X-ray flares has introduced some difficulties
to the standard fireball model and renewed the interest in baryon free 
fireballs (see \S\S~\ref{sec:application-GRB} for further discussion). 

\section{Basic Jet Physics}
Let $\rho^\prime$, $p^\prime$, $e^\prime$, and $h^\prime=(e^\prime+p^\prime)/\rho^\prime c^2$ denote
the proper rest mass density, total pressure, total energy density, and
dimensionless specific enthalpy of the fluid, respectively (henceforth, prime refers to quantities 
measured in the comoving frame).  
The stress-energy tensor takes the form:
\begin{equation}
T^{\alpha\beta} = \rho^\prime h u^{\alpha}u^{\beta} + p^\prime g^{\alpha\beta}+{1\over4\pi}
(F^{\alpha\sigma}F^\beta_\sigma-{1\over4\pi}g^{\alpha\beta}F^2),
\label{T_M} 
\end{equation}
where $u^{\alpha}$ is the four-velocity, $F_{\mu\nu}=\partial _\mu A_\nu
-\partial _\nu A_\mu$ is the electromagnetic tensor, and $g^{\alpha\beta}$ is the metric 
tensor.  The dynamics of the MHD jet is governed by energy and momentum conservation:   
\begin{equation}
T^{\alpha\beta}_{\ \ ;\beta}=0,
\label{Ep-cons}
\end{equation}
mass conservation:
\begin{equation}
(\rho^\prime u^\alpha)_{;\alpha}=0,
\label{nu}
\end{equation}
and Maxwell's equations.  Under the assumption that the MHD flow is ideal 
(i.e., has infinite electric conductivity) Ohms law yields the additional constraint,
$u^\alpha F_{\alpha\beta}=0$.
On scales of interest to the high-energy emission the gravitational force
can be ignored.  Using Eq. (\ref{T_M}) the energy flux can then be expressed as
\begin{equation}
T^{0k} = \rho^\prime c^2 h^\prime\Gamma u^{k}+S^k,
\label{E-flux}
\end{equation}
where $\Gamma$ denotes the bulk Lorentz factor of the flow, $v^k=u^k/\Gamma$ the 
corresponding 3-velocity, and  $S^k$ the k component of the Poynting flux.  The 
infinite conductivity condition implies that the proper electric field must vanish, viz., 
$\vec{E^\prime}=\vec{E}+\vec{v}\times\vec{B}/c=0$.  If in addition the magnetic 
field is toroidal, viz., $\vec{B}=B_\phi\hat{\phi}$, in which case $\vec{B}\perp \vec{v}_p$,
then the Poynting flux reduces to the simple form $\vec{S}=B^2\vec{v}_p/4\pi$.
For a conical jet this is certainly a good approximation well beyond the light cylinder
$R_L=c/\Omega$, where the ratio of toroidal and poloidal magnetic field components
satisfies $|B_{\phi}/B_p|\simeq R/R_L>>1$, with $R$ being the cylindrical radius.

The total jet power is obtained by integrating $T^{0k}$ over a surface 
$\Sigma$ perpendicular to the jet streamlines.  Assuming a conical jet with an 
opening angle $\theta$ and using Eq. (\ref{E-flux}) gives,
\begin{equation}
L_j=\int_{\Sigma} T^{0k}d\Sigma_k=\Gamma u^r \rho^\prime c^2 h^\prime \pi\theta^2r^2(1+\sigma).
\label{L_j}
\end{equation}  
Here $\sigma=B^{\prime2}/4\pi \rho c^2h$ denotes the ratio of Poynting and kinetic 
energy fluxes, and is related to the comoving Alfv\`en 4-speed through
$u_A^\prime=c\sqrt{\sigma}$ (specifically, $u_A^\prime$ is the component of the Alfv\'en 
4-velocity along the direction of wave propagation, as measured in the jet frame).  
Equation (\ref{Ep-cons}) implies that $L_j$ is conserved.  The mass flux carried by the jet is obtained by integrating Eq. (\ref{nu}):
\begin{equation}
\dot{M}=\rho^\prime u^r\pi\theta^2 r^2.
\label{Mdot}
\end{equation}  
This mass flux should be conserved to a good approximation, except for 
situations in which the rest mass content is significantly
altered by processes involving, e.g., pair creation and annihilation\footnote{This includes 
pair cascade models (e.g., Refs.~\refcite{Blandford/Levinson95,Levinson96})
and neutrino-anti neutrino annihilation in GRB models (e.g., 
Ref~\refcite{Aloy05})} or neutron leakage\cite{Vlahakis03}.  Assuming $\dot{M}$ is conserved
and setting $\sigma=0$ and $h=1$ in Eqs. (\ref{L_j}) and (\ref{Mdot}), it is 
readily seen that the asymptotic Lorentz factor of the jet is limited to,
\begin{equation}
\Gamma_{\infty} \le {L_j\over \dot{M}c^2}.
\end{equation}
In the case of magnetically dominated jets for which $\sigma>>1$, acceleration 
to $\Gamma=L_j/\dot{M}c^2$ requires effective conversion of magnetic into kinetic
energy.  This may be difficult to accomplish through direct conversion in
ideal MHD flows\cite{Li92,Lyutikov06}.  
To illustrate this last point consider a magnetically dominated, conical jet section above the light cylinder, where
the magnetic field is nearly toroidal.   Magnetic flux conservation then implies $v^r B_\phi=u^r B_\phi^\prime\propto r^{-1}$, which in turn implies that in the relativistic limit ($v^r\sim 1$) the Poynting power
$L_B\simeq \Gamma^2v^r (B_\phi^{\prime2}/4\pi)\pi\theta^2r^2$ is to a good approximation
conserved.   Now, because $L_j$ and $L_B$ are conserved so do $L_k=L_j-L_B$ and $\sigma=L_B/L_k$, suggesting that at least on conical streamlines direct conversion of magnetic to kinetic energy is highly inefficient (see e.g., Ref~\refcite{Lyutikov06} for a more detailed treatment).  
Similar behavior has been observed also in self-similar models.  It has been argued that the slow conversion may be a consequence of the magnetic field geometry invoked. Indeed, recent analysis of the trans-field equation
has demonstrated that effective magnetic field conversion can be achieved in the case
of a paraboloidal field\cite{Beskin06}.  Other solutions to the sigma problem 
include magnetic reconnection and dissipation by MHD and plasma instabilities.

Table 1 summarizes the range of absolute jet luminosities, Lorentz factors and variability timescales $\Delta t$ inferred for different classes of high-energy sources.  It is generally expected that certain properties will scale with the mass of the black hole, henceforth measured in units 
of solar mass, $m_{BH}=M_{BH}/M_{\odot}$.  In order to elucidate this scaling we find it convenient
to measure luminosities in units of the Eddington luminosity, $L_{Edd}=10^{38}m_{BH}$ erg s$^{-1}$,
and distance in units of the Schwarzchild radius, $r_g=3\times10^{5} m_{BH}$ cm. 
We denote by ${\cal L}_j=L_j/L_{Edd}$ the dimensionless jet power and by $\tilde{r}=r/r_g$ 
the dimensionless radius.  A canonical choice of parameters is 
${\cal L}_j\sim 1$, $ m_{BH}\sim10^8$ for AGNs; ${\cal L}_j\sim 1$, 
$ m_{BH}\sim 3$ for microquasars; ${\cal L}_j\sim 10^{12}$, $ m_{BH}\sim 3$ for GRBs;
and  ${\cal L}_j\sim 10^{8}$, $m_{NS}\sim1$ for giant flare events in magnetars.
In terms of these parameters the proper jet energy density is given by
\begin{equation}
u_j^\prime={L_j\over c\pi(\Gamma\theta)^2r^2}=10^{16}{{\cal L}_j
\over m_{BH}\tilde{r}^2(\Gamma\theta)^2}\qquad{\rm erg\ cm^{-3}},
\end{equation}
and the proper magnetic field by
\begin{equation}
B^\prime={2(\xi_B L_j/c)^{1/2}\over (\theta \Gamma)r}=4\times 10^8
\left(\xi_B {\cal L}_j\over m_{BH}\right)^{1/2}\tilde{r}^{-1}(\theta\Gamma)^{-1}\qquad {\rm Gauss},
\label{B}
\end{equation}
where $\xi_B=L_B/L_j=\sigma/(1+\sigma)$.  The comoving baryon density can be expressed 
in terms of $\eta_p$, the fraction of the jet energy carried by non-relativistic baryons, as
\begin{equation}
n_b^\prime =10^{19}{\eta_p{\cal L}_j
\over m_{BH}\tilde{r}^2(\Gamma\theta)^2}\qquad{\rm cm^{-3}}.
\label{n_b}
\end{equation}

\begin{table}[ph]
\tbl{Characteristic parameters of relativistic jet sources.}
{\begin{tabular}{@{}cccc@{}} \toprule
    & $L_j$ & $\Gamma$ & $\Delta t$ \\
 & (erg/s) &  &  \\ \colrule
GRB\hphantom{00} & \hphantom{0}10$^{47}$-10$^{50}$ & \hphantom{0}10$^{2}-10^3$ & millisec - min. \\
AGN\hphantom{00} & \hphantom{0}10$^{42}$-10$^{47}$ & \hphantom{0} 5 - 50 & hours - years \\
MQ\hphantom{00} &\hphantom{0}10$^{37}$-10$^{40}$ &\hphantom{0}1 - 10 &\hphantom{0}days \\
GF\hphantom{00} & \hphantom{0}10$^{43}$-10$^{46}$ & \hphantom{0}1 &\hphantom{0}seconds \\ \botrule
\end{tabular} \label{ta1}}
\end{table}

\subsection{Energy Dissipation}
\label{sec:jet-diss}
Dissipation of the bulk energy occurs on various scales and by different 
mechanisms.  The rapid variability of the $\gamma$-ray emission observed in
blazars and GRBs suggests that a considerable fraction of the bulk energy 
dissipates already at rather small radii, $10^2-10^6 r_g$.  The dissipation mechanism 
may involve internal shocks, collimation shocks, and/or
magnetic field instabilities.   The latter possibility is particularly relevant
to magnetically dominated outflows in which dissipation by MHD shocks is 
inefficient.  The conversion of magnetic to kinetic energy in Poynting flux dominated jets 
is a long standing issue.  Models of shocks with magnetic 
dissipation have been developed and used to calculate lightcurves 
in blazars.\cite{Romanova92}  However, the physics of magnetic field
dissipation has not been addressed in those models.  Attempts to study particular mechanisms,
including dissipation driven by magnetic reconnection, kink and toroidal field instabilities 
are presented in e.g., Refs.~\refcite{Li92,Sikora05}.    

Quite generally, the radius at which waves created by a fluctuating
source steepen into shocks is $r_{diss}\sim\Gamma^2\gamma_A^2 c\delta t$,
where $\gamma_A$ is the Lorentz factor associated with disturbance speed (e.g., 
the Alfv\`en speed), as measured 
in the fluid rest frame, and $\delta t$ is the characteristic timescale over which 
the parameters of the injected fluid vary\cite{Levinson/putten97}.  Since 
$\delta t \simgt r_g/c$, we expect dissipation of the bulk energy to occur at radii
$\tilde{r}_{diss}>\Gamma^2$ or,
\begin{equation}
r_{diss} >3\times10^5 m_{BH}\Gamma^2\qquad {\rm cm}.
\label{r_diss}
\end{equation}

\subsection{Blast-Waves, Cocoons and Afterglow Emission}
Additional dissipation occurs at radii $10^9-10^{11} r_g$ behind so-called external shocks
that result from collision of the jet with the ambient medium (see Fig. \ref{fig:disk-ext}).  
The radio lobes observed in radio galaxies (FR2 sources in particular) and microquasars, 
and the afterglow emission in GRBs are clear signatures of the associated blast waves.     
In GRBs and other type II supernovae the jet must first break out of the dense envelope of the progenitor star.
This drives a radiation mediated shock that propagates outwards through the stellar envelope.
Under certain conditions high-energy photons and VHE neutrinos 
may be produced during the shock breakout phase\cite{Waxman/Meszaros01}.  

The dynamics of the blast wave and the properties of the cocoon emission depend on the parameters of the ambient medium.  In situations of impulsive energy injection as in, e.g., GRBs and other core collapse supernovae, the evolution of the system at sufficiently late times may become self-similar.  In the ultra-relativistic case one can show that during the self-similar phase the Lorentz factor of an adiabatic blast wave declines with observer time as $\Gamma\propto t_{obs}^{-3/8}$ (see, e.g., Ref.~\refcite{Blandford76}).  This results in a power law decay of the late time afterglow flux in GRBs.
Observations confirm that, at least in some cases, the self-similar model is a good description of the 
blast wave dynamics (see Ref.~\refcite{Waxman06} and references therein).  However, the constraints imposed on the parameters of the shocked plasma are puzzling.  Specifically, the strength of the post shock magnetic field inferred from observations is a factor $\sim10^8$ larger than the typical interstellar magnetic field\cite{Waxman06}, suggesting that the post shock field is somehow generated in the shock and reaches MHD scales before decaying.  Moreover, as mentioned in \S\S~\ref{sec:obs-GRB} and discussed further in \S\S~\ref{sec:application-GRB} recent observations of the early afterglow emission are inconsistent with the predictions of the self-similar afterglow model.

\subsection{Particle Acceleration}
\label{sec-part-acc}

The collisionless shocks created in the jet, and the shocks driven by the jet into the 
surrounding medium, provide sites for particle acceleration.
Both electrons and protons accelerate at the shock front.  The mechanism commonly invoked
to explain the inferred energy distribution of the non-thermal particles is first order 
Fermi acceleration \cite{Blandford/Eichler87}.  This mechanism naturally produces 
a power law distribution of particles, $d n/d\epsilon\propto \epsilon^{-p}$, with the exponent 
$p$ depending on the shock compression ratio.  For adiabatic shocks, test particle theory predicts
$p=2$ in non-relativistic shocks \cite{Blandford/Eichler87} and $p\simeq2.2$
in relativistic shocks \cite{Keshet/Waxman05}, in remarkable agreement with the 
observations.  Energy losses and nonlinear effects may be important under some circumstances 
and can alter the resultant spectrum.
The diffusive shock acceleration theory does not address the so-called injection problem,
namely what fraction of the thermal particles is injected into the Fermi process. 
The observations indicate that an appreciable fraction of the bulk 
energy is tapped for the acceleration of particles to non-thermal energies, suggesting efficient
redistribution inside the shock transition layer.  
The mechanism responsible for the scattering of particles upstream and downstream is yet 
another open issue.  The scattering is most likely anomalous, involving collective 
plasma instabilities that are poorly understood at present.  Further progress in our 
understanding of the microphysics involved requires detailed plasma simulations of collisionless shocks.
Nonetheless, some basic results that are directly relevant to the high-energy emission can
be derived in a simple manner, as we now show. 

The maximum comoving energy of accelerated particles is 
determined by equating he acceleration time with the relevant loss time. 
The time it takes to accelerate a particle to energy  $\epsilon^\prime$
can be expressed as $t_{acc}=\eta r_L/c $, where $r_L=\epsilon^\prime/eB^\prime$
is the corresponding Larmor radius, and $\eta\simlt1$ is an efficiency 
factor that depends on detailed physics.  
For protons radiative losses are typically small, and the relevant loss time is the escape time 
from the system, which in the jet rest frame is given roughly by $t_{esc}=r\theta/c\Gamma$.  
Equating $t_{acc}$ and $t_{esc}$, using Eq. (\ref{B}) and adopting $\eta=1$ yields a maximum 
comoving proton energy,
\begin{equation}
\epsilon^{\prime}_{p,max}\le \epsilon_{conf}=\theta eB^\prime r\simeq
3.6\times10^{16}(\xi_B{\cal L}_{j}m_{BH})^{1/2}\Gamma^{-1}\qquad{\rm eV}. 
\label{eps_max} 
\end{equation}
As shown in \S~\ref{sec:photopion} below, at sufficiently small radii the maximum proton energy may be  
limited by photohadronic losses rather than escape and may be smaller than the confinement limit 
given by the last equation.  
  
The electron spectrum injected at the shock front is likely to be modified by 
radiative cooling, and conceivably pair cascades in sufficiently compact regions
\cite{Blandford/Levinson95}. 
The maximum electron energy is determined by equating the acceleration rate $t_{acc}^{-1}$
with the total cooling rate.  Including both synchrotron 
and inverse Compton cooling, and using again Eq. (\ref{B}), the upper cutoff energy 
of the electron spectrum can be expressed as:
\begin{equation}
\epsilon_{e,max}\simeq10^{9.5}\left({\xi_B{\cal L}_j\over m_{BH}}\right)^{-1/4}
\left(1+{u_{ph}^\prime\over u_B^\prime}\right)^{-1/2} (\theta\Gamma\tilde{r})^{1/2}\qquad {\rm eV},
\label{eps_e,max}
\end{equation}
where $u_{ph}^\prime$ denotes the comoving energy density of the soft radiation field inside the jet.

The energy of thermal electrons depends on the shock velocity and the rate at which 
energy is transfered from ions to electrons.  The mean energy per proton behind the shock
is $\sim(\Gamma_s-1)m_pc^2$, where $\Gamma_s$ is the shock Lorentz factor as measured in the 
frame of the upstream fluid.  The mean electron energy is some fraction of the latter, which
we parametrize as\footnote{with this parametrization $\xi_e=1$ defines equipartition for $\Gamma_s-1\simeq1$} 
\begin{equation}
\epsilon_{e,th}\sim \xi_e m_pc^2/2.
\label{epsilon_eth}
\end{equation}
The thermal electrons just behind the shock will
loose their energy quickly if the synchrotron cooling time, 
$t_{syn}\simeq 6\pi m_e^2c^3/\sigma_T\epsilon_{e,th} B^{\prime2}$, is much shorter than the 
dynamical time $t_d=r/c\Gamma$.  Rapid cooling of thermal electrons 
(i.e., $t_{syn}/t_d<<1$) is expected at radii
\begin{equation}
\tilde{r}<10^{6.5} {\xi_e \xi_B{\cal L}_j\over\theta^2\Gamma^3}.
\end{equation}
Adopting ${\cal L}_j=1$, $\theta^2\Gamma^3\sim1-10$ 
for AGNs and microquasars, we conclude that fast cooling occurs at radii 
$\tilde{r}\simlt (10^{5.5}-10^{6.5})\xi_e\xi_B\sim10^{4}-10^5$ in those sources.  In GRBs, for which
${\cal L}_j=10^{12}$, fast cooling of electrons is expected on even larger scales.

The evolution of the electron (and proton) spectrum is governed by a set of 
kinetic equations that incorporate particle injection and acceleration, 
particle creation and annihilation, and relevant energy loss processes
(see e.g., Refs.~\refcite{Mastichiadis/Kirk97,Mastichiadis/Kirk95}).  A complete treatment of the
problem is rather complicated and requires detailed plasma simulations. Some 
parametrization of the microphysics is commonly invoked in order to simplify 
the analysis.  The simplest approach is to assume that the electrons are, on the average,
continuously injected and re-accelerated in a dissipation 
layer of width $\Delta\simlt r$.  If the cooling length is shorter than 
$\Delta$ a roughly steady electron spectrum will be formed in the
dissipation layer.  Assuming impulsive injection and ignoring pair cascades,
the electron energy distribution below the thermal peak ($\epsilon_e<\epsilon_{e,th}$)
reads
\begin{equation}
{dn_e(r,\gamma_e)\over d\gamma_e}=\kappa(r)\gamma_e^{-2},
\label{dn_e/dg}
\end{equation}
where $\gamma_e=\epsilon_e/m_ec^2$, and $\kappa(r)$ is 
some fraction $\chi_e$ of the total electron density of the shocked plasma, that depends 
roughly on the ratio of the cooling length and $\Delta$.
If the jet composition is dominated by electron-proton plasma, then 
the electron density behind the shock can be estimated using 
Eq. (\ref{L_j}), whereby we obtain (adopting $\sigma<1$ in the dissipation region)
\begin{equation}
\kappa(r)\simeq {\chi_e\Gamma_su^\prime_j\over m_pc^2}=
10^{19}{\chi_e\Gamma_s{\cal L}_j\over m_{BH}}(\theta\Gamma\tilde{r})^{-2}\qquad{\rm cm^{-3}}.
\label{n_e(r)}
\end{equation}
The factor $\Gamma_s$ accounts for compression behind the shock.
The energy distribution above the peak depends on the injected spectrum. 
Assuming efficient injection with a spectrum $Q_{inj}\propto\gamma_e^{-q}$
that joins smoothly the thermal peak, the electron distribution 
above the peak can be written as 
$dn_e(r,\gamma_e)/d\gamma_e=n_e(r)\gamma_{e,th}^{p-2}\gamma_e^{-p}$;  
$\gamma_{e,th}<\gamma_e\le \gamma_{max}$, with $p=q+1$.

\section{Radiation Processes}
\subsection{Low Energy Emission and Target Radiation Fields}
\subsubsection{Synchrotron emission}

\begin{figure}[pb]
\centerline{\psfig{file=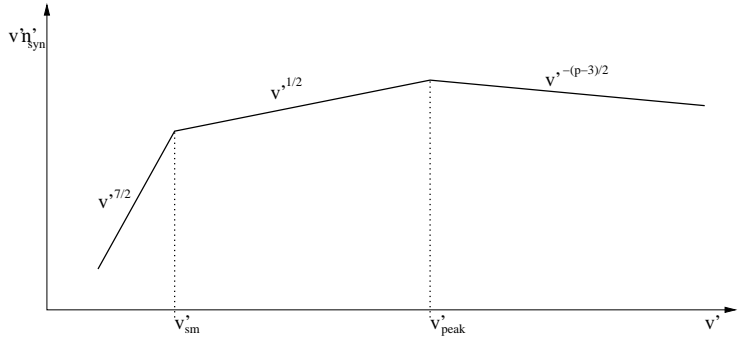,width=9.7cm}}
\vspace*{8pt}
\caption{The local spectral energy distribution of synchrotron photons, $h\nu^\prime n_{syn}^\prime$. The peak frequency $\nu_{peak}^\prime$ and the turnover frequency $\nu_{sm}^\prime$ are indicated.  \label{fig-sync}}
\end{figure}

The local synchrotron spectrum can be well represented by a broken power
law, characterized by several break energies that depend on radius.
At sufficiently low frequencies the synchrotron source
becomes self-absorbed.  The turnover frequency $\nu_{sm}^\prime$ at which the 
synchrotron absorption depth is unity, depends 
on the electron density at the corresponding energy.  As shown below,
the turnover frequency is typically smaller than the peak frequency 
produced by the thermal electrons.  Adopting for illustration the electron distribution
given by Eqs. (\ref{dn_e/dg}) and (\ref{n_e(r)}), which 
is appropriate in the fast cooling regime, one finds
\begin{equation}
h\nu_{sm}^\prime\simeq3\times10^{3}\left({\chi_e\xi_B{\cal L}_j^2\over m_{BH}}\right)^{1/3}
(\theta\Gamma)^{-4/3}\tilde{r}^{-1} \qquad{\rm eV}.
\end{equation}
The comoving spectral energy distribution of synchrotron photons peaks at a frequency 
$\nu_{peak}^\prime=\gamma^2_{e,th}\nu_g^\prime$, where $\gamma_{e,th}=\epsilon_{e,th}/m_ec^2$,
with $\epsilon_{e,th}$ given by Eq. (\ref{epsilon_eth}), is the random Lorentz
factor of the thermal electrons, and $2\pi\nu_g^\prime=eB^\prime/m_ec$.  By employing 
Eqs. (\ref{B}) and (\ref{epsilon_eth}) we arrive at:
\begin{equation}
h\nu_{peak}^\prime\simeq4\times10^{6}\left({\xi_B{\cal L}_j\over m_{BH}}\right)^{1/2}
\xi_e^2(\theta\Gamma\tilde{r})^{-1}\qquad {\rm eV}.
\label{nu_peak}
\end{equation}
Note that the ratio $\nu_{sm}^\prime/\nu_{peak}^\prime\propto\xi_e^{-2}({\cal L}_jm_{BH}/\xi_B)^{1/6}$ 
is independent of radius and depends only weakly on the other parameters except $\xi_e$.  It is therefore 
generally expected that $\nu_{sm}^\prime<<\nu_{peak}^\prime$ for $\xi_e>0.1$.  Note further
that for GRBs, with $m_{BH}=3$, ${\xi_B\cal L}_j=10^{11}$, $\xi_e=0.3$, $\Gamma=10^2\Gamma_{2}$, 
$\theta=0.1\theta_{-1}$, and $r=\Gamma^2 c \Delta t$ adopted, where $\Delta t=\Delta t_{-3}$
milliseconds being the observed variability time of the prompt emission, the observed peak 
energy is $h\nu_{peak}=\Gamma h\nu_{peak}^\prime\simeq 2$ 
MeV $(\theta_{-1}\Gamma^2_{2}\Delta t_{-3})^{-1}$.  This is consistent with the observed
peak of the prompt GRB spectrum, although large scatter is expected for reasonable
variations of the source parameters.

The density of synchrotron photons per logarithmic frequency interval, $n^\prime_{syn}(r,\nu^\prime)
=dn^\prime_{syn}(r,\nu^\prime)/d\ln\nu^\prime$, is shown schematically in Fig.~\ref{fig-sync}.  Above the turnover frequency $n^\prime_{syn}(r,\nu^\prime)\propto(\nu^\prime/\nu_{peak}^\prime)^{-\alpha}$, with $\alpha=1/2$
below the peak and $\alpha=(p-1)/2\simgt 1$ above the peak. In terms of the fraction
$\xi_{syn}=u_{syn}^\prime/u_j^\prime$, where $u_{syn}^\prime=\int{h\nu^\prime 
n^\prime_{syn}d\ln\nu^\prime}$ is the comoving energy density of synchrotron photons, we have
\begin{equation}
n^\prime_{syn}(r,\nu^\prime)=10^{21}{\xi_{syn}\over A\Gamma\theta
\tilde{r}}\left({{\cal L}_j\over \xi_B m_{BH}}\right)^{1/2}
\left({\nu^\prime\over\nu^\prime_{peak}}\right)^{-\alpha}\ \ {\rm cm^{-3}},
\label{n_syn}
\end{equation}
where $A=2-2(\nu_{sm}^\prime/\nu_{peak}^\prime)^{1/2}+
2[(\nu_{max}^\prime/\nu_{peak}^\prime)^{(3-p)/2}-1]/(3-p)$
is a normalization factor.  It should be emphasized that for an inhomogeneous 
source the observed spectrum may differ from the intrinsic local 
spectrum.  In particular,
for an unresolved conical jet the observed spectrum should appear 
flat, $S_\nu\propto\nu^0$, (see, e.g., Ref. \refcite{Blandford87}).

The comoving density of synchrotron photons is constrained by the observations.
For a source at a redshift $z$ observed at viewing angle $\theta_n$ (defined as 
the angle between the jet axis and the sight line), the observed peak frequency is 
$\nu_{peak}=\delta\nu^\prime_{peak}/(1+z)$, where $\delta=[\Gamma(1-\beta\cos\theta_n)]^{-1}$
is the associated Doppler factor.  The flux density at $\nu_{peak}$, as measured on Earh, 
can be expressed as
\begin{equation}
S_\nu={hcn^\prime_{syn}(r_{peak})\over4}\left({r_{peak}\theta\over d_L}\right)^2
\delta^{k},
\label{S_nu}
\end{equation}
where $d_L$ is the luminosity distance, and $r_{peak}$ is the radius from
which photons at the peak originated, and is related to $\nu_{peak}^\prime$
through Eq. (\ref{nu_peak}). The index $k$ depends on whether the photosphere
is at rest with respect to the observer (continuous jet), in which case 
$k=2$, or whether it is moving (emitting blob) in which case $k=3$.  Eq. (\ref{S_nu}) is 
of little use if the source is unresolved, which is usually the case at scales
relevant to the high-energy emission.  However, the rapid variability observed
during intense flare states can be 
used as additional constraint on the source size.  The latter is related to the 
observed variability time $\Delta t$ via 
$r_{peak}\simlt c\Delta t\Gamma\delta/(1+z)$.  As seen, the comoving density of synchrotron photons inferred
from the observed flux density is sensitive to the assumed Doppler factor;
$n^\prime_{syn}\propto S_\nu\delta^{-(2+k)}$.  

\subsubsection{Disk radiation}
An important source of ambient photons is the accretion disk surrounding the compact
object.  Denoting by ${\cal L}_d=L_d/L_{Edd}$ the dimensionless luminosity of the accretion 
disk radiation, and assuming blackbody emission from the inner disk regions yields
a rough estimate for the source temperature: $T_d\sim10^7({\cal L}_d/m_{BH})^{1/4}$ k. 
Consequently, the $\nu F_\nu$ spectrum of the accretion disk radiation should peak in the 
soft X-ray band in stellar X-ray sources, and in the UV band in 
AGNs, in accord with observations.  However, the observations seem to imply the presence
of additional components that contribute to the spectrum of the central continuum source. 
In particular, a power law extension of the peak emission up to hard X-ray/soft $\gamma$-ray
energies is typically seen in both quasars and Galactic X-ray sources.  This component is 
commonly attributed to a hot, tenuous corona above the disk that scatters disk photons.

In Quasars the observed spectral energy distribution of the central continuum source is characterized 
by a big blue bump (BBB), with a peak energy that lies in the range 
$h\nu_{BB}\sim20 -50$ eV.  Above the BBB the spectrum can be 
approximated as a power law with index $s=1.5$ up to an energy of about 0.1-0.5 keV 
\cite{Laor90,Laor94}, although it may slightly 
vary from source to source.  At higher energies the spectrum flattens and has a slop of $s\sim 0.5$ up 
to the upper cutoff energy at about a few hundred keV.  Below the BBB the spectrum flattens again to 
$s<1$.  A fraction $\xi_d(r)$ of the disk radiation is intercepted by the jet at radius $r$.  This 
includes both direct illumination and reprocessed emission, as shown schematically in Fig 
\ref{fig:disk-ext} (see \S \ref{sec:application} for further discussion).  
The associated target photon spectrum (photon density per logarithmic frequency interval) adopted 
below for blazars has the form
\begin{equation}
n_{ext}(r,\nu)\simeq 6\times10^{25}{\xi_d{\cal L}_d\over m_{BH}\tilde{r}^2}\left({h\nu_{BB}\over 
{\rm 25\ eV}}\right)^{-1}\left({\nu\over\nu_{BB}}\right)^{-s}
\left[1+\left({\nu\over \nu_X}\right)^{s-1/2}\right]\ {\rm cm^{-3}},
\label{n_ext}
\end{equation}
with $s=1.5$ for $\nu>\nu_{BB}$ and $s=0.7$ for $\nu<\nu_{BB}$.

The spectrum of the central continuum source is not as well constrained in microquasars.
Although extension of the thermal component to a few hundred keV is typically seen in those objects, it is not clear at present whether it is produced in the jet or in the disk corona.  For our calculations below we shall adopt Eq. (\ref{n_ext}) with $h\nu_{BB}=250$ eV for the target photon density contributed by the external radiation source in microquasars.
We emphasize that the pair-production and photopion opacities at most energies are determined by the portion of the target photon spectrum near and bellow the peak, particularly in microquasars. The seed photons above the peak interact mainly with the lower energy quanta.  It is therefore anticipated that the results presented below wont be sensitive to the shape of the target photon spectrum well above the peak.

\subsubsection{Stellar radiation field}
In microquasars associated with HMXBs, e.g., Cygnus X-1, LS 5039, LS I +61 303,
the companion star, typically O or B spectral type, has a luminosity
$L_\star\sim10^{38} - 10^{39}$ ergs s$^{-1}$, and a spectrum that can be well
approximated by a black body spectrum with a temperature $T_\star\sim40000$ k.  
The binary separation in those systems is of the order of $a\simeq10^{12}$ cm.  At jet radii $r\simlt a$ the stellar photons incident into the jet at relatively large angles and are blue shifted in the jet frame. 
It is therefore naively expected that the external target photon field in the jet will be dominated by stellar radiation at jet radii $r>R(\xi_d L_d/L_\star)^{1/2}$, where $R^2=a^2+r^2$ (see Fig. \ref{fig:MQSW}).
However, the Thomson optical depth of the stellar wind 
may largely exceed unity at jet radii $r\simlt a$ (see Eq. [\ref{tau-MQ-wind}] below), in which case the
jet may be shielded on these scales.  Detailed analysis of Compton scattering of stellar radiation photons in microquasars is 
presented in Ref.~\refcite{Dermer/Botcher06}

\subsection{Inverse Compton Scattering: ERC and SSC emission}
The number density per unit frequency per unit radius of $\gamma$-rays produced by inverse 
Compton scattering of a soft photon distribution $n_s^\prime$ can be expressed as
\begin{equation}
{dn^\prime_{\gamma}(\nu^\prime_\gamma,\mu^\prime_\gamma,r)\over d\nu^\prime_{\gamma}dr}=
\int\sigma^\prime_c(\nu^\prime_s,\mu^\prime_s,\nu^\prime_\gamma,\mu^\prime_\gamma)
n^\prime_s(\nu^\prime_s,\mu^\prime_s,r)d\ln\nu^\prime_s d\mu^\prime_s,
\label{IC-kernel}
\end{equation}
where $\sigma^\prime_c(\nu^\prime_s,\mu^\prime_s,\nu^\prime_\gamma,\mu^\prime_\gamma)$
is a Compton scattering kernel that depends, in general, on the differential cross 
section and the energy distribution of relativistic electrons.
In situations in which the seed radiation field in the jet is dominated by external 
photons (ERC emission) the comoving photon distribution is given by
\begin{equation}
n^\prime_s(\nu^\prime_s,\mu^\prime_s,r)=(\nu^\prime_s/\nu_s)^3n_{ext}(\nu_s,\mu_s,r),
\end{equation}
where $\nu_s=\nu_s^\prime\Gamma(1+\beta\mu_s^\prime)$, $\mu_s=(\mu_s^\prime+\beta)/(1+\beta\mu_s^\prime)$,
with $\beta=(1-\Gamma^{-2})^{1/2}$ being the bulk 3-speed .
For an isotropic power law distribution $n_{ext}(\nu_s,\mu_s,r)=n_0(r)\nu_s^{-\alpha}$;
$\nu_{s,min}<\nu_s<\nu_{s,max}$, the comoving photon distribution is beamed in the backward direction
and can be approximated as (e.g., Ref.~\refcite{Blandford/Levinson95})
\begin{equation}
n^\prime_s(\nu^\prime_s,\mu^\prime_s,r)\simeq{2^{\alpha+2}\Gamma^{\alpha+1}\over(\alpha+1)}
n_0(r)\nu_s^{\prime-\alpha}\delta(1+\mu^\prime_s);\qquad \Gamma\nu_{s,min}<\nu_s^\prime<\Gamma\nu_{s,max}.
\label{n_s-target-ext}
\end{equation}
This can be readily generalized to a broken power law or to the standard soft radiation field given
in Eq. (\ref{n_ext}).  Expressions for the comoving soft-photon density in the case of direct illumination are presented in Ref.~\refcite{Dermer/Schlik02}.  The spectrum of scattered photons can be computed using 
Eqs. (\ref{IC-kernel}) and (\ref{n_s-target-ext})
once the electron distribution is specified.  Redistribution functions of inverse Compton scattering
are presented in Ref.~\refcite{Blandford/Levinson95} for the standard soft spectrum adopted above.
Equation (\ref{eps_e,max}) suggests that $\sim10$ GeV and $\sim1$ TeV photons can be produced in microquasars 
and blazars, respectively, even at the smallest radii (but they will be quickly absorbed if produced below 
the $\gamma$-sphere).  From Eq.~(\ref{epsilon_eth}) we further conclude that
in flat spectrum radio quasars and in microquasars thermal electrons scatter disk photons near the black body peak 
to observed energies $h\nu_\gamma\sim10^6\Gamma^2\xi_e^2h\nu_{BB}\sim 0.1 - 10$ MeV.  This may be the origin of the
MeV bumps seen in those objects.  Further discussion on the applications of ERC models to individual sources 
is given in \S~\ref{sec:application}.

Synchrotron-self Compton (SSC) emission refers to situations whereby the synchrotron photons are Compton scattered by the relativistic electrons accelerated in situ\cite{Konigl81}\cdash\cite{Konopelko03}.  In one-zone SSC models the 
same electron population producing the synchrotron photons scatter them to higher energies.  
Such models have been used to infer the parameters of the emission zone in TeV blazars
(e.g., Ref. \refcite{Mastichiadis/Kirk97}).  As mentioned in \S~\ref{sec:obs-blazars}, the basic assumption 
is that the synchrotron emission gives rise to the low energy component of the observed SED, while
inverse Compton scattering of the synchrotron photons produce the high-energy component.
The observed cutoff frequencies $\nu_s$ and $\nu_c$ of the low and high energy components, respectively,
can be used to relate the magnetic field and maximum electron energy $\gamma_{e,max}m_ec^2$ to the 
Doppler factor $\delta$.  For $\nu_s$ we have
\begin{equation}
\delta\gamma_{e,max}^2{eB\over 2\pi m_ec}=\nu_{s}.
\end{equation}
Photons having frequency $\nu_c$ are produced by IC scattering in the Klein-Nishina regime
when 
\begin{equation}
(\nu_{s}\nu_{c})^{1/2}>{m_ec^2\over h}\simeq 10^{20}\qquad {\rm Hz}.
\label{nu_s}
\end{equation}
This is typically the case in the TeV BL Lac objects.  In that case we have
\begin{equation}
\delta\gamma_{e,max}m_ec^2=h\nu_{c}.
\label{nu_c-KN}
\end{equation}
In sources for which the scattering is in the Thomson regime $(\nu_s\nu_c)^{1/2}<<10^{20}$ Hz,
condition (\ref{nu_c-KN}) should be replaced by 
\begin{equation}
\gamma^2_{e,max}\nu_{s}=\nu_{c}.
\end{equation}
For the TeV blazars $\nu_s\sim10^{18}$ Hz and $\nu_c\sim10^{27}$ Hz during quiescent states.  Eqs. 
(\ref{nu_s}) and (\ref{nu_c-KN}) can be solved to yield
\begin{eqnarray}
\gamma_{e,max}=10^{6.5}\delta^{-1}(\nu_c/10^{27}\ {\rm Hz}),\label{g_max,ssc}\\
B=5\times10^{-3}\delta (\nu_c/10^{27}\ {\rm Hz})^{-2}(\nu_s/10^{18}\ {\rm Hz})\qquad {\rm Gauss}.
\label{B,ssc}
\end{eqnarray}
In cases where the Doppler factor can be estimated the parameters of the emission region are fully 
determined (see \S\S \ref{sec:application-BLLAc}).

\subsection{Pair Creation and Annihilation}
At sufficiently small radii the {\em compactness parameter}, 
$l_\gamma=L_\gamma\sigma_T/(m_ec^2r)\simeq10^4{\cal L}_\gamma\tilde{r}^{-1}$, largely exceeds
unity for all classes of sources considered here.  It is, therefore, expected that
$\gamma$-rays will not be able to escape from the inner jet regions without creating pairs.
Moreover, as will be shown below the bulk outflow energy cannot be carried by e$^\pm$ pairs
at small radii, owing to rapid pair annihilation.  This implies that the energy released by the central
engine must be transported to larger radii by baryons and/or magnetic fields.  
In what follows we consider $\gamma$-ray absorption and pair annihilation in some
detail.

\subsubsection{The $\gamma$-sphere}

\begin{figure}[pb]
\centerline{\psfig{file=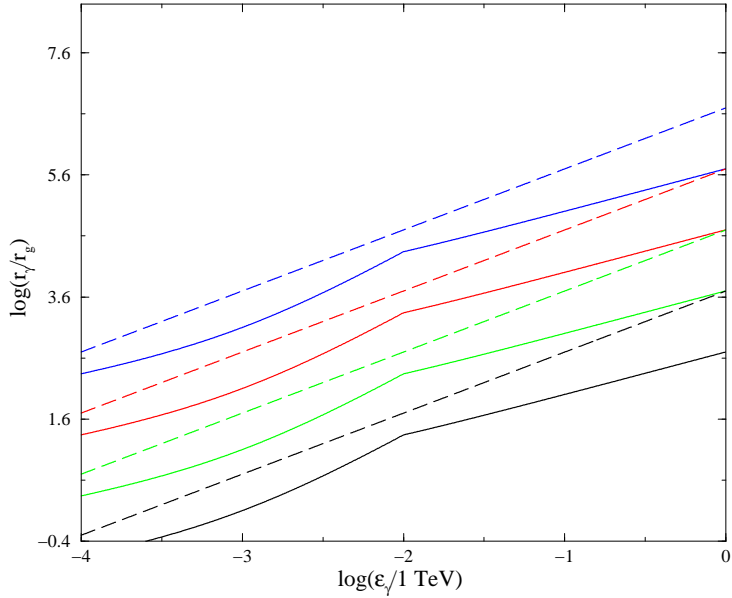,width=9.7cm}}
\vspace*{8pt}
\caption{Dimensionless $\gamma$-spheric radius $\tilde{r}_{\gamma}=r_\gamma/r_g$ 
versus $\gamma$-ray energy $\epsilon_\gamma$.  Solid lines show 
$\tilde{r}^{ext}_\gamma$ for $\nu_{BB}/\nu_X=0.1$ and $\xi_d{\cal L}_d(h\nu/25 {\rm eV})^{-1}=10^{-2}$ (blue), 
$10^{-3}$ (red), $10^{-4}$ (green) and $10^{-5}$ (black).  The dashed lines 
show $\tilde{r}^{syn}_\gamma$ for $({\xi_e^2\xi_{syn}{\cal L}_j/A\theta^2\Gamma^3\delta})=10^{-2}$ 
(blue), $10^{-3}$ (red), $10^{-4}$ (green) and $10^{-5}$ (black).
\label{fig-tau}}
\end{figure}

$\gamma$-rays produced in the jet can combine with soft photons to produce electron-
positron pairs.  The pair production cross section is given by                                        
\begin{equation}              
\sigma_{\gamma\gamma}={3\over16}\sigma_T(1-\beta^2)   
\left[(3-\beta^4)\ln\left({1+\beta\over1-\beta}\right)-2\beta(2-\beta^2)\right],                                    
\label{sigma_p}                                                                                                 
\end{equation}                     
where $\beta$ is the speed of the electron and the positron in                                                        
the center of momentum frame\cite{Gould/Shruer67}, and                                                         
is related to the $\gamma$-ray energy $\epsilon^\prime_\gamma$, the soft (target) photon 
energy $h\nu^\prime_s$ and the cosine of the angle between their momenta, $\mu$, through 
\begin{equation}                             
(1-\beta^2)={2m_e^2c^4\over(1-\mu)h\nu^\prime_s\epsilon^\prime_\gamma};\qquad0\le\beta<1;                        
\end{equation}    
$\beta=0,\mu=-1$ corresponds to the threshold for pair production that we express as
\begin{equation}
\epsilon_{\gamma,th}^\prime\simeq 2.5\times10^{11}
\left({h\nu_s^\prime\over {\rm 1 eV}}\right)^{-1}\qquad {\rm eV}.
\label{thrs}
\end{equation}
As a rough estimate one can adopt $\sigma_{\gamma\gamma}\sim 0.2\sigma_T$ near threshold.
For a comoving soft photon distribution $n^\prime_s(\mu,\nu^\prime,r)$ (per unit volume, 
per $\ln\nu^\prime$, per steradian) the pair production opacity 
at some radius $r$ is
\begin{equation}
\kappa^\prime_{\gamma\gamma}(\epsilon^\prime_\gamma,r)=2\pi \int_{-1}^1d\mu(1-\mu) 
\int_{\ln[2/(1-\mu)\epsilon^\prime_\gamma]}{d\ln \nu^\prime_s}
n^\prime_{s}(r,\mu,\nu^\prime_s)\sigma_{\gamma\gamma}(\epsilon_\gamma^\prime,\nu^\prime_s,\mu).
\label{kappa_gg} 
\end{equation}
Both the external radiation intercepted by the jet and the synchrotron photons produced inside 
the jet provide an opacity to pair production.  The corresponding optical depth  
can then be expressed as $\tau_{\gamma\gamma}=\tau_{\gamma\gamma}^{ext}+\tau_{\gamma\gamma}^{syn}$
where $\tau_{\gamma\gamma}^{ext}$ denotes the contribution of external photons and
$\tau_{\gamma\gamma}^{syn}$ the contribution of synchrotron photons. 

Consider first the contribution of the synchrotron radiation field.  
Denoting by $\epsilon^\prime_{\gamma,peak}$ the threshold $\gamma$-ray energy 
for pair creation by interaction with synchrotron peak photons, and using
Eqs. (\ref{nu_peak}) and (\ref{thrs}) one finds 
\begin{equation}
\epsilon^\prime_{\gamma,peak}=6\times10^{4}\left({\xi_{B}{\cal L}_j\over m_{BH}}\right)^{-1/2}
\xi_e^{-2}\theta\Gamma\tilde{r}\qquad{\rm eV}.
\label{epsilon_gpeak}
\end{equation}
Assuming that the photon distribution is isotropic in the jet frame, and substituting 
$n_s^\prime=n_{syn}^\prime$ in Eq. (\ref{kappa_gg}), where $n_{syn}^\prime$ is the 
distribution of synchrotron photons given by Eq. (\ref{n_syn}), we obtain
\begin{equation}
\tau^{syn}_{\gamma\gamma}(\epsilon_\gamma^\prime,r)=
\int_r^\infty{\kappa^\prime_{\gamma\gamma}(\epsilon_\gamma^\prime,y){dy\over\Gamma}}\simeq
30\left({\xi_{syn}\over A\theta\Gamma^2}\right)\left({{\cal L}_jm_{BH}\over\xi_B}\right)^{1/2}
\left({\epsilon^\prime_{\gamma}\over\epsilon^\prime_{\gamma,peak}}\right)^{\alpha},
\label{tau^syn}
\end{equation}
with $\alpha=1/2$ for $\epsilon^\prime_{\gamma}>\epsilon^\prime_{\gamma,peak}$ and $\alpha=(p-1)/2\simeq1$ 
for $\epsilon^\prime_{\gamma}<\epsilon^\prime_{\gamma,peak}$.
For AGNs, GRBs and magnetars we anticipate $\tau_{\gamma\gamma}^{syn}>>1$ at $
\epsilon^\prime_\gamma=\epsilon^\prime_{\gamma,peak}$, while for microquasars $\tau_{\gamma\gamma}^{syn}\sim1$.
This implies that at any given radius only $\gamma$-rays having energies $\epsilon^\prime_{\gamma}
<\epsilon^\prime_{\gamma,peak}$ can escape the system.  From Eq. (\ref{tau^syn}) it is readily seen 
that the  '$\gamma$-sphere', defined as the radius 
$r^{syn}_{\gamma}(\epsilon_{\gamma})$ beyond which the pair production optical depth to infinity
is unity, viz., $\tau^{syn}_{\gamma\gamma}(r^{syn}_\gamma,\epsilon^\prime_\gamma)=1$, is 
proportional to $\epsilon^\prime_{\gamma}$.  In terms of the observed $\gamma$-ray 
energy $\epsilon_{\gamma}=\delta\epsilon_{\gamma}^\prime$ we have
\begin{equation}
\tilde{r}^{syn}_{\gamma}=5\times10^{6}\left({\xi_e^2\xi_{syn}{\cal L}_j
\over A\theta^2\Gamma^3\delta}\right)
\left({\epsilon_{\gamma}\over{\rm 10\ GeV}}\right),
\label{r^syn_gamma}
\end{equation}
where Eq. (\ref{epsilon_gpeak}) and (\ref{tau^syn}) have been used with $\alpha=1$.
In the case of GRBs we conclude from Eq. (\ref{r^syn_gamma}) that $\gamma$-ray absorption
by pair production may be important at radii $\tilde{r}\simlt10^{7}$ even at MeV energies
(see e.g., Ref.~\refcite{Peer04} for detailed calculations).

To compute the optical depth to pair production on external photons, we transform 
Eq. (\ref{kappa_gg}) to the star frame and take $n_s=n_{ext}$. We further assume that
the external radiation field is roughly isotropic, which is a good approximation in
situations whereby the target radiation field is dominated by reprocessed or scattered 
disk photons.  With these simplifications the corresponding optical depth reads
\begin{equation}
\tau^{ext}_{\gamma\gamma}\simeq2\times10^6\left({\xi_d{\cal L}_d\over \tilde{r}}\right)
\left({h\nu_{BB}\over{\rm 25\ eV}}\right)^{-1}
\left({\epsilon_{\gamma}\over\epsilon_{\gamma,BB}}\right)^s\left[1+\left({\epsilon_{\gamma}
\over\epsilon_{\gamma,x}}\right)^{1/2-s}\right],
\label{tau^ext}
\end{equation}
where $\epsilon_{\gamma,BB}={\rm 10 GeV} (h\nu_{BB}/{\rm 25\ eV})^{-1}$ and 
$\epsilon_{\gamma,x}=\epsilon_{\gamma,BB}(\nu_{BB}/\nu_{x})$.  The associated 
$\gamma$-spheric radius can be readily obtained from Eq. (\ref{tau^ext}) 
and the equation $\tau^{ext}_{\gamma\gamma}(r,\epsilon_{\gamma})=1$.  The 
$\gamma$-spheric radii $r_\gamma^{syn}$ and
$r_\gamma^{ext}$ are exhibited in Fig. \ref{fig-tau}

\subsubsection{Pair annihilation}  
As argued in \S\S~\ref{sec-part-acc} in the inner jet regions the cooling time of electrons and positrons is much shorter than the dynamical time and, therefore, all pairs are expected to quickly cool to subrelativistic energy.   
The comoving pair density at a given radius, $n^\prime_\pm(r)$, is then limited by electron-positron annihilation\cite{Blandford/Levinson95}.  Equating the annihilation time, $t^\prime_{ann}\sim(\sigma_{ann} n_{\pm}^\prime v_{\pm})^{-1}$, where $v_{\pm}$ denotes the thermal velocity of pairs in the jet frame, with the outflow time $t^\prime_f\sim r/c\Gamma$ gives $n_\pm\simeq \Gamma c/\sigma_{ann} v_\pm r$. 
The maximum power that can be carried by pairs at radius $r$ is then 
\begin{equation}
L_\pm\simeq(n_+^\prime+n_-^\prime)\Gamma^2m_ec^3\pi\theta^2r^2\simeq2\Gamma^3m_ec^3\pi\theta^2r/\sigma_{ann}.
\end{equation}
Consequently, at radii smaller than the {\it annihilation radius},
\begin{equation}
\tilde{r}_{ann}\simeq10^3{{\cal L}_j\over \Gamma^3\theta^2},
\label{r_ann}
\end{equation}
$L_\pm < L_j$ and so the jet power cannot be carried by pairs.  For powerful blazars and microquasars this means that some alternative carrier of energy and momentum must be present at radii $\tilde{r}\simlt 10^2 -10^3$.  For GRBs Eq. (\ref{r_ann}) gives $\tilde{r}_{ann}\sim10^9$.  We note, however, that the above analysis does not hold in compact regions where the comoving pair temperature exceeds $\sim1$ MeV, e.g., the base of a GRB fireball (see Ref.~\refcite{Paczynsky86}).  In this case kinetic equilibrium is established whereby pair creation balances pair annihilation.  However, as the pair outflow accelerates the comoving temperature quickly drops owing to adiabatic cooling and the pairs will eventually annihilate.  This typically happens at radii $\tilde{r}\sim10^2-10^3$, well below $\tilde{r}_{ann}$.

\subsection{Inhomogeneous Pair Cascade Models}
\label{sec:pair-cascades}
Injection of electrons and positrons to energies well above the local threshold $\gamma$-ray
energy would lead to a development of intense pair cascades.  In the inhomogeneous
pair cascade model\cite{Blandford/Levinson95}  the pairs are injected with high energy                            
over many octaves of jet radius. At a given radius, $\gamma$-rays are produced through 
inverse Compton scattering of external disk photons.
The energy of a freshly emitted $\gamma$-ray                                           
is then degraded, via pair cascades, into many $\gamma$-rays                                                         
with energy near the threshold energy below which the                                                       
$\gamma$-rays can escape to infinity without being further converted into                                           
pairs. This threshold energy increases with radius, as seen in Fig. \ref{fig-tau}, so that       
higher energy $\gamma$-rays tend to originate from larger radius.   

For sufficiently flat electron injection spectra, the resultant $\gamma$-ray spectrum
reflects the spatial profiles of energy dissipation and target photon intensity,
because the asymptotic flux is dominated by $\gamma$-rays that emerged from near to 
their associated $\gamma$-spheres.   
For $\kappa(r)\propto r^{-p}$ in Eq. (\ref{dn_e/dg}) and $\xi_d\propto r^{-q}$ 
in Eq. (\ref{n_ext}) the emitted $\gamma$-ray spectrum is roughly a power law
with spectral index $\alpha_\gamma\sim\tilde{\alpha} p/(1+q)$, where $\tilde{\alpha}$ 
is the average spectral index of the target photon field\cite{Blandford/Levinson95}.


\subsection{Constraints from Variability}
Constraints on the source parameters can be derived from the variability
of the high-energy emission.  Suppose that a variability timescale $\Delta t$
has been measured at some observed $\gamma$-ray energy $\epsilon_\gamma$.
This implies  that the emission at the observed energy originated from radii
$r\simlt r_{var}\equiv\Gamma\delta c \Delta t/(1+z)$, where $\delta$ is the Doppler factor
and $z$ is the redshift of the source. 
At the emission zone the pair production optical depth at the observed energy 
must not exceed unity, viz., $\tau^{syn}_{\gamma\gamma}(r_{var},\epsilon_\gamma/\delta)<1$,
implying $r^{syn}_{\gamma}(\epsilon_\gamma)< r_{var}$.  Suppose now that the 
synchrotron flux near the peak can be measured simultaneously.  Using 
Eq. (\ref{S_nu}) with $k=3$ and the fact that $r_{peak}\le r_{var}$ we deduce that 
\begin{equation}
n^\prime_{syn}(\nu_{peak})\ge{4 d_L^2 S_\nu\over h c r^2_{var}\theta^2\delta^3}.
\label{n_syn_const}
\end{equation}
For a low redshift source ($z<1$) we take $d_L\simeq10^{28}h^{-1}z$ cm.
By combining Eqs. (\ref{n_syn}), (\ref{r^syn_gamma}) and (\ref{n_syn_const}),
and adopting $A=5$ we arrive at
\begin{equation}
(\xi_e^2\xi_{syn}/\xi_B)^{1/2}\delta^{5}\simgt2\times10^{11}\Delta 
t_h^{-3/2}(\Gamma\theta)^{-2}z^2(\epsilon_\gamma/{\rm 1 TeV})^{1/2}S_{{\rm Jy}},
\label{Var-const}
\end{equation}
where $S_{\rm Jy}$ is the flux at $\nu_{peak}$ in Janskys and $\Delta t_h$ is
$\Delta t$ measured in hours.

\section{Hadronic Processes}
Relativistic jets are also potential sources of ultra-high energy cosmic rays (UHECR)
and VHE neutrinos.  This of course requires that 
a considerable fraction of the bulk energy will be carried by baryons.
In contrast to electromagnetic emission that can have either leptonic or hadronic
origin, VHE neutrino emission is a unique diagnostic of a hadronic content. 
High-energy neutrinos can be produced in astrophysical jets mainly through the 
decay of charged pions: 
\begin{eqnarray}
\pi^{-}\rightarrow \mu^-+\bar{\nu}_\mu\rightarrow e^-
+\bar{\nu}_e +\nu_\mu+\bar{\nu}_\mu ,\nonumber\\
\pi^{+}\rightarrow \mu^++\nu_\mu \rightarrow e^++\nu_e +\nu_\mu+\bar{\nu}_\mu.
\end{eqnarray}
Decay of neutral pions,
\begin{equation}
\pi^{0} \rightarrow \gamma +\gamma,
\end{equation}
leads to production of high energy photons, and under certain conditions 
may compete with inverse Compton scattering.  The pions may be produced 
through collisions of protons with target photons ($p\gamma$)\footnote{In cases whereby the
neutron lifetime is longer than the loss time due to photomeson interactions the 
production rate of $\pi^-$ via ($n\gamma$) collisions becomes comparable to the $\pi^+$ 
production rate.}, or via inelastic ($pp$) and ($pn$) collisions.  The latter mechanism produces
both $\pi^+$ and $\pi^-$ in roughly equal numbers, whereas the former mechanism
produces mainly $\pi^+$ in regions of moderate optical depth.
Therefore, it may be possible to discriminate between the two production 
modes of charged pions by measuring flavor composition (e.g., Refs. \refcite{Learned95}--\refcite{Kashti05}),
or even to probe new physics\cite{Learned95,Athar00},
although the interpretation of such measurements may not be straightforward\cite{Kashti05}.

\subsection{UHECR Production}
As explained in \S~\ref{sec-part-acc} a substantial fraction of the energy 
dissipated behind shocks can, in principle, be taped for acceleration of protons to ultrahigh energies 
with a power law spectrum $dn^\prime_p/d\epsilon^\prime_p\propto\epsilon_p^{\prime -2.2}$ 
and an upper cutoff $\epsilon^\prime_{p,max}$ that is limited by confinement 
(see Eq. \ref{eps_max}).  Consequently, production of UHECRs of observed energy 
$\epsilon_p=\Gamma\epsilon_p^\prime$ requires
\begin{equation}
\xi_B m_{BH}{\cal L}_j\ge10^8(\epsilon_p/10^{20.5}{\rm eV})^2,
\label{Hillas}
\end{equation} 
which essentially leaves GRBs with $m_{BH}\sim3$, $\xi_B\simeq0.1$, ${\cal L}_j\simeq 10^{12}$,
and powerful blazars with $m_{BH}\sim10^9$, $\xi_B\simeq0.1$, ${\cal L}_j\simeq 1$
as the main candidates for astrophysical UHECRs sources\footnote{
Condition (\ref{Hillas}) does not hold in systems that violate the ideal MHD condition, e.g., vacuum gaps in 
dormant AGNs\cite{Boldt99,Levinson00} and magnetars.  In this case the luminosity is limited only by the rate at which 
UHECRs are produced.}.  Whether protons can actually be accelerated by 
relativistic shocks to the highest energies observed is yet another issue.  

\subsection{Inelastic Nuclear Collisions} 
Pion production via interactions of the ultra-high energy nuclei accelerated in the 
jet and the cold jet material is typically inefficient in sub-Eddington sources, but may be relevant
in highly super-Eddington sources.  For the baryon density given in Eq. (\ref{n_b})
the optical depth for inelastic nuclear collisions is approximately
\begin{equation}
\tau_{pp}\simeq \sigma_{pp}n_b^\prime{r\over \Gamma}\simeq10^{-1}{\eta_p{\cal L}_j\over \theta^2\Gamma^3\tilde{r}},
\end{equation}
where $\sigma_{pp}=50$ mbn has been adopted, which is suitable for rough astrophysical 
estimates.  For Eddington sources  like AGNs and microquasars, for which ${\cal L}_j\le1$, this is 
always much less than unity.  In GRB models that invoke $\eta_p\simeq1$ at 
$\Gamma\sim100$, $\theta\Gamma\sim10$ and ${\cal L}_j\sim10^{12}$ this process may 
be relevant at radii $\tilde{r}\simlt10^7$.

\subsubsection{Target ions from stellar winds in microquasars}
In microquasars with high-mass stellar companions a potentially important source 
of target protons is provided by the stellar wind\cite{Romero03}\cdash\cite{Torres/Halzen06},
as illustrated in Fig. \ref{fig:MQSW}.  
Typical mass loss rates and terminal velocities of O stars are of the order of 
$\dot{M}_w\sim10^{-5} M_{\odot} {\rm yr^{-1}}$ and $v_w\sim 2000$ km s$^{-1}$, 
respectively\cite{Lamers/Cassinelli}.  At a distance $R$ from the 
companion star the wind density is $n_p\simeq \dot{M}_w/(4\pi m_p R^2 v_w)$, and the optical 
depth for pp collisions is
\begin{equation}
\tau_{pp}\simeq \sigma_{pp}n_p R\simeq10^{2.5}\left({\dot{M}_w\over 10^{-5}M_{\odot}{\rm yr^{-1}}}\right)
\left({R\over 10^{2}R_{\odot}}\right)^{-1}\left({v_w\over 10^{2}\ {\rm km\  s^{-1}}}\right)^{-1}.
\label{tau-MQ-wind}
\end{equation}
If stellar wind ions can penetrate the jet then effective production of pions in the jet/wind interaction site,
and consequent emission of VHE neutrinos and $\gamma$-rays is anticipated.  Such an origin has been proposed for the TeV emission from the two Tev microquasars detected thus far.  Kelvin-Helmholtz instabilities
at the interface between the jet and slow wind may give rise to entrainment of the wind material.  Whether  
significant fraction of the stellar wind ions can be intercepted by the jet is yet an open issue.  Owing to rotation of the binary system modulation of the observed fluxes is expected.

\subsubsection{Neutron pick up in GRBs}
In some scenarios, the ultra-relativistic GRB-producing jet is envisaged
to be ensheathed by a slow, baryon rich wind emanating from the
disk surrounding the black hole.  This wind contains free neutrons
out to a radius of $\sim10^{9}-10^{11}$ cm, that can diffuse across magnetic 
field lines into the central baryon free jet.  The leaking neutrons are then
picked-up and converted to protons in a collision avalanche\cite{Levinson/Eichler03}.
It has been suggested\cite{Levinson/Eichler03} that baryon loading of GRB fireballs is accomplished by this 
process; the number of captured neutrons has been found to be in reasonable agreement
with existing limits on the GRB baryonic component.
The charged decay and collision products of the neutrons become ultra-relativistic immediately, and
a VHE neutrino burst is produced with an efficiency that can
exceed 0.5. Other signatures may include lithium, beryllium and/or
boron lines in the supernova remnants associated with GRB's and
high polarization of the $\gamma$-rays\cite{Levinson/Eichler04}.  

\subsection{Photomeson Production}
\label{sec:photopion}
Collision of a proton with a photon can lead to neutrino and $\gamma$-ray production
through the following reactions:
\begin{eqnarray}
p+\gamma\rightarrow n+\pi^{+}, \\
p+\gamma\rightarrow p+\pi^{0}\nonumber.
\end{eqnarray}
The photopion production cross section peaks at $\sigma_{p\gamma}\sim 0.5$ mb at the 
$\Delta$ resonance and drops to $\sigma_{p\gamma}\sim0.1$ mb at higher energies 
where multipion production dominates.  The inelasticity is $K_\pi\sim0.2$ near the 
threshold and $K_\pi\sim0.5$  in the multipion production regime. 
The threshold proton energy for which head-on collision with a target photon
is at the $\Delta$ resonance is
\begin{eqnarray}
\epsilon_{p,th}^\prime\simeq 7\times10^{16}\left({h\nu_s^\prime\over {\rm 1 eV}}\right)^{-1}\ \ {\rm eV},
\label{eps_pthrs}
\end{eqnarray}
where $h\nu^\prime_s$ is the energy of the target photon, as measured in the jet rest frame.

Consider now photomeson interactions on the target synchrotron photons produced in the jet.
For illustration we adopt the synchrotron spectrum given in Eq. (\ref{n_syn}). 
The comoving proton energy for which interaction 
with photons near the peak is at the $\Delta$ resonance is related to the corresponding
pair production threshold through Eqs. (\ref{thrs}), (\ref{epsilon_gpeak}) and (\ref{eps_pthrs}):
\begin{equation}
\epsilon^\prime_{p,peak}=10^{5.5}\epsilon^\prime_{\gamma,peak}=2\times10^{10}
\left({\xi_{B}{\cal L}_j\over m_{BH}}\right)^{-1/2}
\xi_e^{-2}\theta\Gamma\tilde{r}\ \ {\rm eV}.
\label{eps_ppeak}
\end{equation}
The energy loss rate of protons due to photomeson interactions is $t_{p\gamma}^{-1}
\sim\sigma_{p\gamma}K_\pi n^\prime_{syn}c$.  Equating the latter with the acceleration rate
$t_{acc}^{-1}\sim eB^\prime c/\epsilon_p^\prime$, using Eq. (\ref{n_syn}) with $\alpha=1$ and
Eq. (\ref{eps_ppeak}), and assuming single pion inelasticity factor of $K_\pi=0.2$ 
one obtains an upper limit on the proton energy:
\begin{equation}
\epsilon_{p,max}^\prime\simlt 2\times10^{15}\left({A\over \xi_{syn}}\right)^{2/3}
\left({{\cal L}_j\over m_{BH}}\right)^{-1/6}\xi_e^{-2/3}\xi_B^{1/2}
(\theta\Gamma\tilde{r})^{1/3}\qquad {\rm eV}.
\label{eps_max2}
\end{equation}
\begin{figure}[pb]
\centerline{\psfig{file=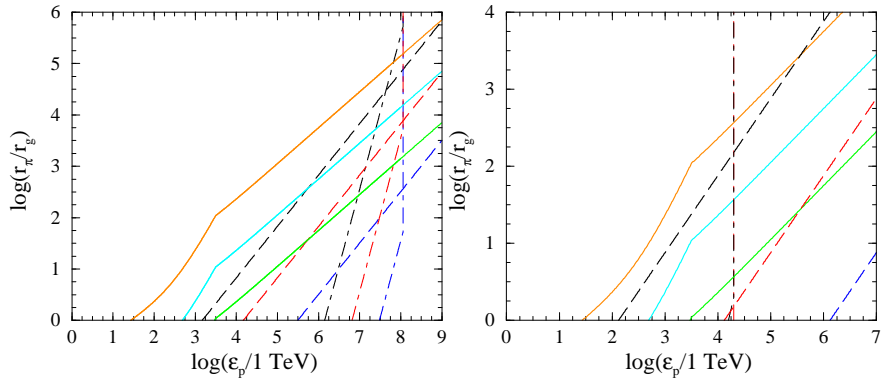,width=11.7cm}}
\vspace*{8pt}
\caption{Dimensionless $\pi$-spheric radius $\tilde{r}_{\pi}=r_\pi/r_g$ versus 
proton energy measured by an on-axis observer $\epsilon_p=\Gamma\epsilon_p^\prime$.  
The left panel corresponds to a choice of parameters that represents a 
typical blazar like 3c279: $\theta\Gamma=1$, $\Gamma=10$, $m_{BH}=10^8$, and
the right panel to a choice of parameters representing a typical microquasar like GRS 1915: 
$\theta=0.1$, $\Gamma=3$, $m_{BH}=3$. 
The solid lines in both windows show 
$\tilde{r}^{ext}_\pi$ for  $\xi_d{\cal L}_d(h\nu/25 {\rm eV})^{-1}=10^{-2}$ (orange), 
$10^{-3}$ (cyan), and $10^{-4}$ (green).  
The dashed lines show $\tilde{r}^{syn}_\pi$ for 
${\cal L}_j=1$, $\xi_B=0.1$, $\xi_e=0.3$ and $\xi_{syn}/A=10^{-1}$ (black), $10^{-2}$ (red)
and $10^{-3}$ (blue).  The dotted-dashed lines correspond to the maximum proton 
energy $\epsilon_{p,max}=\Gamma\epsilon_{p,max}^\prime$ computed using the minimum of 
Eqs. (\ref{eps_max}) and (\ref{eps_max2}), with the same color code employed 
for the dashed curves.  The break in the dotted-dashed lines corresponds to the transition from radii  
at which the maximum proton energy is limited by inelastic collisions 
to radii at which it is limited by escape (see Eq. [\ref{transition}]).
\label{fig-rpi}}
\end{figure}
Comparing the latter limit with Eq. (\ref{eps_max}) then implies that the maximum
proton energy is limited by photomeson losses at radii 
\begin{equation}
\tilde{r}\simlt 6\times10^3\left({\xi_e\xi_{syn}{\cal L}_j\over A}\right)^2
\left({m_{BH}\over \theta\Gamma^4}\right),
\label{transition}
\end{equation}
and by confinement at larger radii. From Eqs. (\ref{eps_max}), (\ref{eps_ppeak})
and (\ref{eps_max2}) it is evident that protons can be accelerated to energies
exceeding that required for photomeson interactions with synchrotron peak photons, 
viz., $\epsilon^\prime_{p,peak}\le\epsilon^\prime_{p,max}$, at radii
\begin{equation}
\tilde{r}\simlt 2\times10^6\left({\xi_{B}{\cal L}_j\over\theta\Gamma^2}\right)
{\rm min}\left\{1\ ,\ {15\xi_e^2A\Gamma\over \xi_{syn}}
\left({\xi_B\over m_{BH}{\cal L}_j}\right)^{1/2}\right\}.
\end{equation}
A rough estimate of the photopion optical depth contributed by synchrotron photons gives
\begin{equation}
\tau^{syn}_{p\gamma}(\epsilon_p^\prime,r)\simeq\sigma_{p\gamma}n^\prime_{syn}{r\over\Gamma}=
0.15\left({\xi_{syn}\over A\theta\Gamma^2}\right)\left({{\cal L}_jm_{BH}\over\xi_B}\right)^{1/2}
\left({\epsilon^\prime_{p}\over\epsilon^\prime_{p,peak}}\right)^{\alpha},
\label{tau^syn_pg}
\end{equation}
with $\alpha=1/2$ for $\epsilon^\prime_p>\epsilon^\prime_{p,peak}$ and 
$\alpha\simeq1$ for $\epsilon^\prime_p<\epsilon^\prime_{p,peak}$. 
As seen the photopion opacity increases with 
increasing proton energy and is largest at $\epsilon_{p,max}^\prime$.
This trend is expected for any reasonable synchrotron spectrum, at least up 
to a proton energy at which interactions at the $\Delta$ resonance involves synchrotron photons
having energies above the self-absorption frequency $h\nu_{sm}^\prime$.
In analogy to the $\gamma$-sphere we introduce the {\em $\pi$-sphere}, the radius 
$r_\pi^{syn}$ that solves the equation $\tau^{syn}_{p\gamma}(\epsilon^\prime_{p},r_\pi^{syn})=1$. 
In terms of the transformed energy $\epsilon_p=\Gamma\epsilon_p^\prime$ we obtain from Eq. (\ref{tau^syn_pg})
\begin{equation}
\tilde{r}_\pi^{syn}= 50\xi_e^2\left({0.15\xi_{syn}\over A}\right)^{1/\alpha}
\left({{\cal L}_j\over \theta^2\Gamma^4}\right)^{(1+\alpha)/2\alpha}
\left({m_{BH}\over \xi_B}\right)^{(1-\alpha)/2\alpha}
\left({\epsilon_p\over {\rm 1\ TeV}}\right),
\label{r_pi}
\end{equation}
where $\alpha=1$ for $0.15({\xi_{syn}/ A\theta\Gamma^2})({{\cal L}_jm_{BH}/\xi_B})^{1/2}>1$
and $\alpha=1/2$ when the opposite inequality holds. 

\begin{figure}[pb]
\centerline{\psfig{file=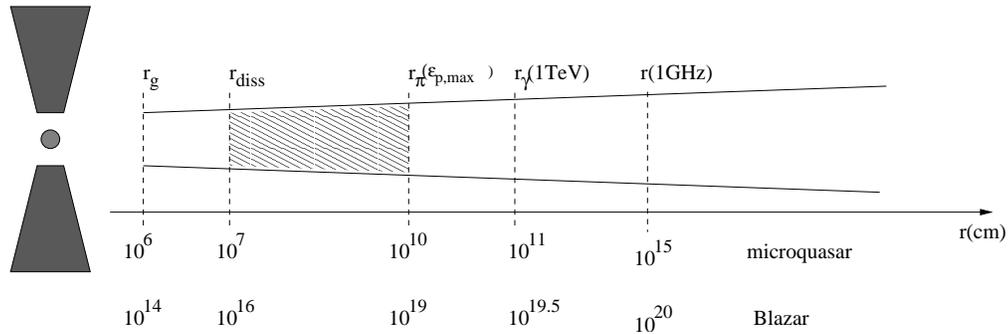,width=14.7cm}}
\vspace*{8pt}
\caption{Characteristic scales exhibited for typical Galactic and extragalactic sources. 
The radius above which shocks can form $r_{diss}$, the $\pi$-sphere at maximum 
proton energy $r_\pi(\epsilon_{p,max})$, the $\gamma$-sphere of a TeV photon $r_\gamma$(1 TeV), and the 
radio photosphere at 1 Ghz $r$(1Ghz) are indicated.  The shaded
area corresponds to the region where effective production of VHE neutrinos can occur.}
\label{fig-scales}
\end{figure}

As for pair production, external photons intercepted by the jet also provide
targets for $p\gamma$ collisions.  The corresponding optical depth, computed
using the external radiation spectrum given in Eq. (\ref{n_ext}), reads:
\begin{equation}
\tau^{ext}_{p\gamma}\simeq 10^4\left({\xi_d{\cal L}_d\over \tilde{r}}\right)
\left({h\nu_{BB}\over{\rm 25\ eV}}\right)^{-1}
\left({\epsilon_{p}\over\epsilon_{p,BB}}\right)^s\left[1+\left({\epsilon_{p}
\over\epsilon_{p,x}}\right)^{1/2-s}\right],
\label{tau^ext_pg}
\end{equation}
where $\epsilon_{p,BB}={\rm 10^{15.5} eV} (h\nu_{BB}/{\rm 25\ eV})^{-1}$
and  $\epsilon_{p,x}=\epsilon_{p,BB}(\nu_{BB}/\nu_x)$.  The $\pi$-spheric radius
$r_\pi^{ext}(\epsilon_p)$ associated with the external radiation field can be readily 
obtained by setting $\tau_{p\gamma}^{ext}=1$ in the 
last equation.  The radii $\tilde{r}_\pi^{syn}$ and $\tilde{r}_\pi^{ext}$ are plotted 
in figure \ref{fig-rpi}.  A comparison of the different scales derived above is presented in Fig. \ref{fig-scales}

\subsection{Radiative Cooling of Pions and Muons}
\label{sec:pi-cool}
Radiative cooling of charged pions and muons can limit the energy of 
the resultant neutrinos and, more importantly, the overall efficiency of neutrino production.
For a pion having energy $\epsilon^{\prime}_{\pi}=m_{\pi}c^2\gamma^\prime_{\pi}$, the ratio of 
synchrotron cooling time and decay time, $\tau_{\pi}=2\times10^{-8}\gamma^\prime_{\pi}$ s, is given by
\begin{equation}
\frac{t_{syn}}{\tau_{\pi}}=5\times10^{23}\gamma_{\pi}^{\prime -2} B^{\prime-2}
=3\times10^6\gamma_{\pi}^{\prime -2}\left({\xi_B{\cal L}_j\over 
m_{BH}}\right)^{-1}(\theta\Gamma\tilde{r})^2,
\label{t_syn}
\end{equation} 
where $B$ is given in Gaussian units.  From the last equation it is evident that
high energy neutrino emission is severely suppressed in regions where $B\simgt10^{12}$ G,
such as expected near the surface of pulsars and magnetars.  The only exception is 
the gaps formed in charge starved regions in the magnetosphere, where the pions can 
accelerate along magnetic field lines by a parallel electric field\cite{Zhang01}.  
In general pions having energies 
\begin{equation}
\epsilon_{\pi}^\prime > 2.4\times 10^{11}\left({\xi_B{\cal L}_j\over m_{BH}}\right)^{-1/2}
\theta\Gamma\tilde{r} \ \ {\rm eV}
\label{eps_pi}
\end{equation}
will cool radiatively before decaying.  For muons this limit is smaller
by a factor $(\tau_\mu/\tau_\pi)^{1/2}=10$, where $\tau_\mu$ is the muon lifetime.
  
Suppose now that protons can be accelerated to the maximum limit allowed by confinement,
and consider the implications for synchrotron cooling of charged pions produced 
by photomeson interactions of those maximum energy protons.   
To elucidate the limit on the maximum proton energy 
above which the resultant photopions lose their energy radiatively, we 
express equation (\ref{t_syn}) in terms of the confinement energy given in Eq. (\ref{eps_max}).
Using Eq. (\ref{eps_pi}) and adopting inelasticity of $K_\pi=0.2$ for single 
photopion production; that is, taking $\epsilon^{\prime}_{\pi}=0.2\epsilon_{conf}$
yields,
\begin{equation}
\frac{t_{syn}}{\tau_{\pi}}\simeq 2\times 10^{-23}\frac{(\theta m_{BH}\tilde{r})^2}
{(\epsilon_{conf}/10^{20}{\rm \ eV})^4}.
\end{equation} 
This suggests that if protons are accelerated to energies in excess of 
$2\times10^{14}(\theta m_{BH}\tilde{r})^{1/2}$ 
eV, then the consequent charged pions will lose most of their energy before decaying.

Pions also lose energy as a result of inverse Compton scattering of synchrotron photons.
At pion energies 
\begin{equation}
\epsilon_\pi^\prime\ge\epsilon_{\pi,c}=5\times10^9\left({\xi_B{\cal L}_j\over m_{BH}}\right)^{-1/2}
\theta\Gamma\tilde{r}\qquad {\rm eV}
\label{eps_pic}
\end{equation}
the scattering of peak synchrotron photons is in the Klein-Nishina regime.  Comparing 
this with Eq. (\ref{eps_ppeak}) 
we find that $\epsilon_{\pi,c}^\prime/\epsilon_{p,peak}^\prime\simeq 0.25\xi_e^2$,
which for $\xi_e\sim1$ equals roughly the inelasticity factor $K_\pi$.  Consequently,
for plasma at rough equipartition pions produced by protons of energy 
$\epsilon_{p}^\prime>\epsilon_{p,peak}^\prime$ 
satisfy $\epsilon_\pi^\prime\ge\epsilon_{\pi,c}$.  The energy loss time due to 
inverse-Compton scattering is then approximately 
\begin{equation}
t_{IC}\sim {3\over8c\sigma_T n^\prime_{syn}}{\epsilon_\pi^\prime h\nu_{peak}^\prime\over
(m_sc^2)^2}=4\times10^{-5}{A\xi_e^2\xi_B\over \xi_{syn}}\gamma^\prime_\pi \qquad s,
\end{equation}
where $\gamma^\prime_\pi=\epsilon^\prime_\pi/m_\pi c^2$, and the ratio of the latter and the 
pion lifetime is 
\begin{equation}
\frac{t_{IC}}{\tau_{\pi}}\simeq 2\times 10^{3}{A\xi_e^2\xi_B\over\xi_{syn}},
\end{equation} 
independent of pion energy.  Since $A\sim$ a few, it is concluded that inverse-Compton 
losses of pions and muons can be ignored provided $\xi_e^2\xi_B/\xi_{syn}>10^{-4}$
and $\xi_e^2\xi_B/\xi_{syn}>10^{-2}$, respectively.

\begin{figure}[pb]
\centerline{\psfig{file=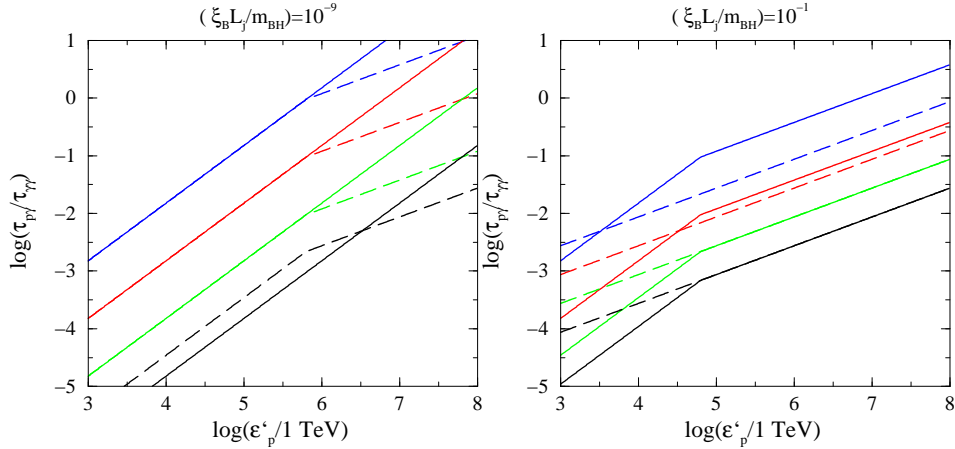,width=12.7cm}}
\vspace*{8pt}
\caption{Opacity ratio $\tau^{syn}_{p\gamma}(\epsilon_p^\prime)/
\tau^{syn}_{\gamma\gamma}(\epsilon_{\gamma}^\prime)$
as a function of comoving proton energy $\epsilon_p{^\prime}$ at $\tilde{r}=10^2$ (dashed lines)
and $10^5$ (solid lines).  The blue, red, green and black colors correspond to comoving $\gamma$-ray
energies of $\epsilon_{\gamma}^\prime=10$, $10^2$, $10^3$, and $10^4$ GeV respectively.
\label{fig-tau-ratio}}
\end{figure}

\begin{figure}[pb]
\centerline{\psfig{file=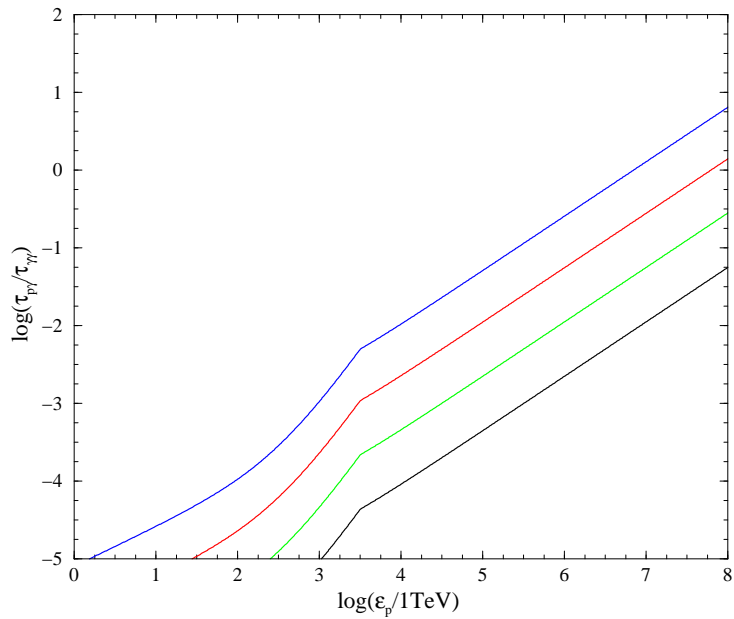,width=9.7cm}}
\vspace*{8pt}
\caption{Opacity ratio $\tau^{ext}_{p\gamma}(\epsilon_p)/
\tau^{ext}_{\gamma\gamma}(\epsilon_{\gamma})$
as a function of observer frame proton energy $\epsilon_p$.  
The blue, red, green and black colors correspond to $\gamma$-ray
energies of $\epsilon_{\gamma}=10$, $10^2$, $10^3$, and $10^4$ GeV respectively.
\label{fig-tau2-ratio}}
\end{figure}

\subsection{Relations Between Pair-Production and Pion-Photoproduction Opacities}
Observations of VHE $\gamma$-rays can constrain the pion photoproduction 
opacity, particularly in situations where rapid variability of 
the VHE $\gamma$-ray flux is observed. 
To illustrate the relationship between the pair production and pion photoproduction
opacities, consider first interactions of $\gamma$-rays and protons with 
the synchrotron spectrum presented in Eq. (\ref{n_syn}).  From Eqs. (\ref{tau^syn}) 
and (\ref{tau^syn_pg}) one finds
\begin{equation}
{\tau_{p\gamma}^{syn}(\epsilon^{\prime}_p,r)\over \tau_{\gamma\gamma}^{syn}(\epsilon^{\prime}_{\gamma},r)}
=\left(\frac{\epsilon^{\prime}_p}{3\times10^5\epsilon^{\prime}_{\gamma}}\right)^{\alpha}
\frac{\sigma_{p\gamma}}{\sigma_{\gamma\gamma}}\simeq4\times10^{-3}
\left(\frac{\epsilon^{\prime}_p}{3\times10^5\epsilon^{\prime}_{\gamma}}\right)^{\alpha},
\label{tau_pg}
\end{equation}
where $\alpha=1/2$ if both $\epsilon_{\gamma}^\prime>\epsilon_{\gamma,peak}^\prime$
and $\epsilon_{p}^\prime>\epsilon_{p,peak}^\prime$ and $\alpha=1$ when the opposite
inequalities hold.  A plot of this ratio in the general case is displayed in 
Fig. \ref{fig-tau-ratio}.  The ratio of opacities associated with the
external radiation field,  $\tau^{ext}_{p\gamma}(\epsilon_p)/
\tau^{ext}_{\gamma\gamma}(\epsilon_{\gamma})$, can be computed in a similar manner and 
is displayed in Fig. \ref{fig-tau2-ratio}.  As seen from the figures, at $\gamma$-ray energies above 
a few TeV the opacity ratio is smaller than unity even at the maximum proton energy.  $\gamma$-ray observations 
at these energies can therefore constrain the neutrino flux that can be emitted from the same region.
A particular application of this result is discussed in \S\S~\ref{sec:application-BLLAc}.

\subsection{Neutrino Flux and Spectrum}

The neutrino flux emitted from the jet depends on several factors:
(i) the fraction of jet energy injected as a power law distribution of 
protons $\xi_p$, (ii) the spectrum of accelerated protons, (iii) 
the cooling times of pions and muons and (iv) the pion production opacity.
It is customary to introduce the parameter $f_\pi(\epsilon_p^\prime)\le1$, 
representing the fraction of proton energy $\epsilon_p^\prime$ 
converted to pions.    In case of pion photoproduction it can be approximated
by $f_\pi(\epsilon_p^\prime)={\rm min}[1,K_\pi\tau_{p\gamma}(\epsilon_p^\prime)]$.
For a proton distribution $n_p(\epsilon_p^\prime)$ the average proton energy
lost to pions can be defined as 
\begin{equation}
\bar{f}_\pi={\int f_\pi(\epsilon_p^\prime)\epsilon_p^\prime n_p(\epsilon_p^\prime)
d\epsilon_p^\prime\over \int \epsilon_p^\prime n_p(\epsilon_p^\prime)
d\epsilon_p^\prime}.
\label{f_pi}
\end{equation}
This average fraction depends on the spectrum of both protons and target photons.
As argued in \S\S~\ref{sec:pi-cool} 
pions are expected to decay prior to significant energy loss, while muons 
may lose significant fraction of their energy before decaying. We shall
therefore conservatively assume that 
a single high energy $\nu_\mu$ (or $\bar\nu_\mu$) is
produced in a single photopion interaction of a proton (or neutron),
corresponding to conversion of $1/8$ of the energy lost to pion
production to muon neutrinos.  In terms of $\xi_p$ and $\bar{f}_\pi$
the flux at Earth of $\nu_\mu$ and $\bar\nu_\mu$ can be expressed as,
\begin{equation}
{\cal F}_{\nu_\mu}\simeq\xi_p\bar{f}_\pi\Gamma^{-1}\delta^3\frac{L_j/8}{4\pi d_L^2}
\simeq10^{-20}\xi_{p}\bar{f}_\pi{\delta^3\over \Gamma}{m_{BH}{\cal L}_{j}\over  d_{L28}^2} 
\qquad {\rm erg\ s^{-1} cm^{-2}},
\label{Nu-flux}
\end{equation}
where $\delta$ is the Doppler factor, and $d_L=10^{28}d_{L28}$~cm is the 
luminosity distance to the source. 

A rough estimate of the neutrino spectrum expected to be produced in sources of 
high optical depth to photopion production can be made using Fig \ref{fig-rpi}. 
Firstly, assuming that neutrinos are produced at radii larger than the 
dissipation radius $r_{diss}$ given by Eq. (\ref{r_diss}) and recalling that
the energy of the resultant neutrino
is $\simeq5$\% of the proton energy, we expect significant $\nu$ flux in the 
energy interval $30$ - $10^5$ TeV in AGNs, $1$ - $10^2$ TeV in microquasars,
and above $10^2$ TeV for GRBs internal shocks\cite{Waxman/Bahcall97}.
Secondly, for a thick target the neutrino spectrum should be similar
to the proton spectrum. Thus, we expect $dn_\nu/d\epsilon_\nu\propto\epsilon_\nu^{-2}$
in the above energy intervals.

The probability that a muon neutrino will
produce a high-energy muon in a terrestrial detector is \cite{Gaisser95}
$P_{\nu\mu}\approx1.3\times10^{-6}E_{\nu,\rm TeV}^{\beta}$, with $\beta=2$
for $E_{\nu}<1$ TeV and $\beta=0.8$ for $E_{\nu}>1$ TeV.  Thus,
for a flat neutrino spectrum above 1~TeV, the
muon flux at the detector is $\approx (P_0/E_0) {\cal F}_{\nu_\mu}$,
where $P_0/E_0=1.3\times10^{-6}{\rm TeV}^{-1}$.  The rate of neutrino induced muon 
detections is
\begin{equation}
\dot N_{\mu}\simeq 2\times10^{-12}\xi_{p}\bar{f}_\pi{\delta^3\over \Gamma}{m_{BH}{\cal L}_{j}
\over d_{L28}^2}(A/1{\rm km}^2)\qquad {\rm day}^{-1},
\label{eq:dotNmu}
\end{equation}
where Eq. (\ref{Nu-flux}) has been used.  Estimates of neutrino yields for each class of objects is given in \S~\ref{sec:application}.

\section{Applications}
\label{sec:application}
\subsection{Flat Spectrum Radio Quasars}
The prodigious emission exhibited by the most powerful $\gamma$-ray blazars 
imply ${\cal L}_j\ge1$ for a black hole mass range $m_{BH}\sim10^8-10^{10}$
inferred for quasars.  Lorentz factors $\Gamma\sim10-30$, as inferred from superluminal
motions, are typical for FSRQs.  Line luminosities and observations of radio quiet
(jetless) quasars suggest disk luminosity of ${\cal L}_d\sim1$.  The main uncertainty
is the fraction $\xi_d$ intercepted by the jet.  As 
seen from Fig. \ref{fig-tau} EGRET (0.1-20 GeV) $\gamma$-spheres are typically located within the broad 
emission line region.  On such scales an important fraction of disk radiation can be 
Thomson scattered across the jet by the broad line gas.  The fraction $\xi_d$ is then
roughly equals the average Thomson depth $<\tau_T>$ of the 
scattering medium\cite{Blandford/Levinson95}.  The standard                                                    
emission line clouds are generally modeled to have free electron column densities of          
$\sim10^{21}$~cm$^{-2}$ and covering factors $\sim0.1$, giving  
$\tau_T>\sim10^{-4}$.  However, the inter-cloud medium, which is                                                   
often invoked as an agent to confine these clouds, is widely supposed to have an 
electron depth a factor $\sim100$ larger (e.g., Ref.~\refcite{Netzer92}).  
Furthermore, bound electrons, which may have a much greater density
than free electrons, will present a roughly Thomson scattering cross section
to hard X-rays.  It is therefore reasonable to expect $\xi_d\sim10^{-2}-10^{-3}$.
Since $({\xi_e^2\xi_{syn}{\cal L}_j/A\theta^2\Gamma^3\delta})<10^{-3}$ is expected for the inferred
values of $\Gamma$ and ${\cal L}_j$, it is concluded 
that the target photon field should be dominated by the reprocessed disk emission 
if $\xi_d\simgt10^{-3}$ (see Fig. \ref{fig-tau}). Detailed models that take into account 
SSC and ERC components\cite{Bottcher99} show that the ratio of ERC and SSC luminosities required by spectral fits 
lies in the range 1 to 10.

The $\gamma$-spheres can be mapped in principle by measuring
temporal variations of the $\gamma$-ray flux in different energy bands during a flare.
If the $\gamma$-ray emission is indeed produced over many octaves of jet radius where 
intense pair cascades at the observed energies are important,
as in the inhomogeneous pair cascade model discussed in \S\S~\ref{sec:pair-cascades}, then we expect the high energy
to either lag the lower energy or to vary more slowly.
With the limited sensitivity and energy band of the EGRET instrument it was practically 
impossible to resolve such effects.  It is hoped that with the upcoming GLAST instrument this will be feasible

As seen from Fig. \ref{fig-rpi}, above the dissipation radius $\tilde{r}_{diss}\simeq\Gamma^2\simeq10^2$,
the photopion optical depth exceeds unity at proton energies above $10^{15}$ eV for  
$\xi_d{\cal L}_d\simeq10^{-2}$.  The maximum proton energy there exceeds $10^{19}$ eV.  
Assuming $n_p(\epsilon_p^\prime)\propto\epsilon_p^{\prime-2}$ we estimate
$\bar{f}_\pi\simeq0.5$.  Likewise, for $\xi_d{\cal L}_d=10^{-3}$ we find $\bar{f}_\pi\simeq0.2$.
With $m_{BH}=10^8$, ${\cal L}_j=1$, $\Gamma=10$, $\xi_p=0.1$ and $\bar{f}_\pi\simeq0.2$ Eq. (\ref{eq:dotNmu})
implies detection rate of about one signal event per year in a cubic-scale neutrino detector.
A similar conclusion has been drawn in Ref.~\refcite{Atoyan01}
who specifically considered the flare observed in 1996 from the quasar 3C279.  

\subsection{TeV Blazars}
\label{sec:application-BLLAc}
As mentioned in \S~\ref{sec:obs-blazars} all known TeV blazars are
BL Lac objects, which are typically much fainter than FSRQs.  The observed bolometric 
luminosity of the TeV blazars during quiescent states is 
of the order of a few times $10^{44}$ ergs s$^{-1}$, with about 10 percents
emitted as VHE $\gamma$-rays. The VHE flux increases by more than a factor
of 10 during flaring states.  In terms of the observed luminosity of VHE photons, denoted henceforth by $L_{VHE}$, 
and the corresponding radiative efficiency $\xi_{rad}$, the jet power can be expressed as 
\begin{equation}
L_{j}=\xi_{rad}^{-1}(\delta^3/\Gamma)^{-1}L_{VHE}.
\label{L_j-BLLac}
\end{equation}

Estimates of the black hole mass in a sample of several dozens BL Lac objects yield a
range of $10^7<m_{BH}<10^{9.5}$, similar to quasars\cite{Woo05}.  In particular, for the 
TeV blazars Mrk 421, Mrk 501 and 1ES 2344+514, Ref.~\refcite{Woo05} estimates
$\log(m_{BH})=8.22\pm0.06$, $8.62\pm0.11$ and $8.74\pm0.18$, respectively.
This implies ${\cal L}_{VHE}=L_{VHE}/L_{Edd}\sim$ $10^{-3}-10^{-2}$.

The absence of emission lines
suggests, as often argued, that the target radiation field in the BL Lac objects
is dominated by the synchrotron photons produced in the jet.  At radio to IR 
frequencies the observed flux lies in the range $10^{-2}-1$ Jy.
Adopting $\Gamma\theta=1$, $S_{Jy}=0.1$, and $z=0.03$ (roughly the redshift measured for Mrk 421 and Mrk 501)
Eq. (\ref{Var-const}) yields
\begin{equation}
\delta\simgt10^{7/5}(\xi_B/\xi_e^2\xi_{syn})^{1/10}\Delta t_h^{-3/10}.
\label{BLLac1}
\end{equation}
As described in \S~\ref{sec:obs-blazars} above, variations of the VHE $\gamma$-ray
flux on timescales $\Delta t_h\simlt1$ have been reported for Mrk 421 and
some other BL Lac objects.  Since quite generally $(\xi_e^2\xi_{syn}/\xi_B)\simlt1$
Eq. (\ref{BLLac1}) implies $\delta\sim 10-50$ during flaring states (see also
Refs.~\refcite{Krawezynski02}--\refcite{Ghisellini05}).
Such high values are in clear disagreement with the much lower values inferred from 
source statistics\cite{Urry91,Hardcastle03} and superluminal motions on parsec scales\cite{Jorstad01}\cdash\cite{Giroletti04}. Various explanations, including jet deceleration\cite{Georganopoulos03,Piner05}, a structure 
consisting of interacting spine and sheath\cite{Ghisellini05}, and opening angle effects\cite{Gopal-Krishna04}
have been proposed in order to resolve this discrepancy.  Under the one-zone SSC model Eqs. (\ref{g_max,ssc}), (\ref{B,ssc}) and (\ref{BLLac1}) give a maximum electron energy $m_ec^2\gamma_{e,max}\sim0.1$ TeV, magnetic field $B\sim0.1$ G, and Doppler factor $\delta\sim30$ in the emission region at a radius of $r\simlt10^{17}$ cm.

Consider now the implications for $\gamma$-ray and neutrino emission.  
First, with the above estimates of the black hole masses
we have $c\Delta t/r_g\sim 1\Delta t_h$.  Since the $\gamma$-spheric radius must
be smaller than the characteristic size of the emission zone we deduce 
that $\tilde{r}_\gamma(\epsilon_\gamma=1TeV)\simlt\delta^2$.
At small viewing angles $\delta\simeq2\Gamma$, implying 
$\tilde{r}_\gamma(\epsilon_\gamma=1{\rm TeV})\simeq \tilde{r}_{diss}$, where 
$r_{diss}$ is the dissipation radius given by Eq. (\ref{r_diss}).  Thus,
the observed variability seems to imply that the unsteadiness of the flow parameters 
occurs over a dynamical timescale. 
Assuming the source is observed at small viewing angles, viz., $\delta\simeq2\Gamma$,
and adopting ${\cal L}_{VHE}=10^{-3}$ (roughly the value measured for Mrk 421 and Mrk 501)
in Eq. (\ref{L_j-BLLac}) gives ${\cal L}_j=10^{-3}\xi_{rad}^{-1}\Gamma^{-2}$.  Substituting the latter result
into Eq. (\ref{eps_e,max}) and taking $\theta\Gamma=1$, we obtain $\epsilon_{e,max}
\simlt10^{14.5}(\xi_{rad}/\xi_B)^{1/4}(\Gamma/10)^{3/2}$ eV, consistent with the observed spectrum.
Likewise, Eq. (\ref{eps_max}) gives $\epsilon_{p,max}<
10^{18.5}(\xi_B/\xi_{rad})^{1/2}(\Gamma/10)^{-1}$ eV.  Combining the latter result with 
Eq. (\ref{eps_ppeak}) we find
$\epsilon_{p,max}^\prime/\epsilon_{p,peak}^\prime\le0.03\xi_e^{2}(\Gamma/10)^{-5}$, which 
combining with Eq. (\ref{BLLac1}) implies $\epsilon_{p,max}^\prime/\epsilon_{p,peak}^\prime<1$.
Since $\tau^{syn}_{\gamma\gamma}(\epsilon_\gamma=1{\rm TeV})<1$ we deduce from Eq. (\ref{tau_pg})
that $\tau^{syn}_{p\gamma}<4\times10^{-2}(\epsilon_p/\epsilon_{p,max})$ and $f_\pi(\epsilon_p^\prime)=
K_\pi\tau^{syn}_{p\gamma}\simeq0.2\tau_{p\gamma}^{syn}<10^{-3}
(\epsilon_p^\prime/\epsilon^\prime_{p,max})$. Substituting $f_\pi$ into Eq. (\ref{f_pi}) 
and assuming $n_p(\epsilon_p^\prime)\propto\epsilon_p^{\prime-2}$ we have $\bar{f}_\pi<
10^{-3}/\ln(\epsilon_{p,max}^\prime/\epsilon_{p,min}^\prime)\simlt10^{-4}$, where 
$\epsilon_{p,min}^\prime\sim$ 1GeV is the lower cutoff energy of the proton energy distribution.
This demonstrates that, contrary to earlier claims, the emitted flux of high energy neutrinos 
should be only a small fraction of the observed VHE $\gamma$-ray flux.  
With the above numbers the rate of neutrino induced muons expected to be measured
by the upcoming detectors is 
\begin{equation}
\dot{N}_\mu<0.03\xi_p(A/1{\rm km^2})\qquad yr^{-1},
\end{equation}
too small to be detected.
 
\subsection{Microquasars}
In microquasars with resolved jets a minimum jet power can be estimated from the observed 
brightness temperature of radio knots.  Such estimates yield $L_j\sim10^{37}-10^{39}$ ergs s$^{-1}$
for most sources\cite{Levinson/Blandford96,Distefano02,Grenier04}.  For systems with a measured 
black hole mass the jet power appears to be close to the Eddington limit, viz., ${\cal L}_j\sim 0.01-1$.
The contributions of the jet, disk and companion star to the observed broad-band
continuum spectrum cannot be directly extracted from the observations.  It is 
believed that in microquasars associated with low-mass X-ray binaries the accretion disk 
is the dominant source of external photons, whereas in microquasars with a massive stellar
companion the stellar radiation makes an important contribution to the external radiation
field\cite{Dermer/Botcher06,Grenier04,Georganopoulos02}.  The observed X-ray emission, whether due to
disk emission or synchrotron emission from the jet implies 
$r_\gamma>>r_{diss}$ at GeV energies, suggesting that, like in blazars, pair cascades 
may be important.  Detailed treatment of $\gamma$-ray emission from leptonic microquasar jets 
and fits to individual sources are given in e.g.,
Refs.~\refcite{Levinson/Blandford96,Atoyan/Aharonian99,Dermer/Botcher06,Georganopoulos02}--\refcite{Bednarek06}.
Hadronic $\gamma$-ray emission has also been considered (e.g., Ref.~\refcite{Romero03,Aharonian/Luis05}).
This is particularly relevant for microquasars with high-mass stellar companions for which, as
can be seen from Eq.(\ref{tau-MQ-wind}), neutral
pions can be effectively produced by inelastic collisions of relativistic protons accelerated
in the jet and the ions in the stellar wind.  In this case the flux of VHE neutrinos should be
comparable to the observed $\gamma$-ray flux.  Neutrino yields have been estimated for the TeV microquasar
LS I +61 303 assuming the observed TeV emission has a hadronic origin\cite{Torres/Halzen06}.  With the upcoming neutrino experiments it would be possible to constrain this model.

Next, we estimate detection rates for VHE neutrinos produced by decay of photopions in intermittent microquasars (see Ref.~\refcite{Levinson/Waxman01} for more details).  The latter exhibit ejection episodes that lasts several days,
with a total energy of $E_j=L_j\Delta t\sim 10^{43}-10^{45}$ ergs.  
The expected number of events in a single burst is
\begin{equation}
N_{\mu}\simeq 20\xi_{p}\bar{f}_\pi{\delta^3\over \Gamma}{E_{j44}
\over d_{22}^2}(A/1{\rm km}^2),
\label{eq:Nmu}
\end{equation}
where $E_{j44}$ is $E_j$ in units of 10$^{44}$ ergs, and $10^{22}d_{22}$ cm is 
the distance to the source.
The duration of the neutrino burst should be of the order of the 
blob's ejection time.  It should precede the associated radio 
outburst, or the emergence of a new superluminal component, that 
originate from larger scales ($\sim 10^{15}$ cm), by several hours.  
For sources directed along our sight line, $\Gamma^{-1}\delta^3\sim 8\Gamma^2$ 
may exceed 100.  Thus, if the fraction 
$\xi_p$ exceeds a few percent, several neutrinos can be detected during
a typical outburst from a source at a distance of a few kpc. 
The typical angular resolution of the planned neutrino telescopes at TeV 
energies (e.g. Ref.~\refcite{Halzen01}) should be $\theta\sim 1$ degree. 
The atmospheric neutrino background flux is 
$\Phi_{\nu,bkg} \sim 10^{-7}\epsilon_{nu,\rm TeV}^{-2.5}/{\rm cm^2 s\,sr}$, 
implying a number of detected background events 
$N_{bkg}\sim 3\times 10^{-2} (\theta /\hbox{deg})^2 t_{\rm day} {\rm km}^{-2}$
per angular resolution element over a burst duration 
$1t_{\rm day}$~day.  The neutrino signals above $\sim10$~TeV 
from a typical micro-quasar outburst should therefore be easily detected above 
the background. 

As a particular example consider the 1994 March 19 outburst observed in GRS 1915+105 \cite{Mirabel94}.
Conservative estimates of the total energy released yield $E_{44}=2-4$\cite{Levinson/Blandford96,Mirabel99}.
From the proper motions measured with the 
VLA, a speed of 0.92c and angle to the line of sight of 
$\theta\simeq70^{\circ}$ are inferred for the ejecta, corresponding to $\Gamma\simeq2.5$ and a 
Doppler factor $\delta\simeq0.58$.  For these values Eq. (\ref{r_diss}) yields a dissipation radius 
of $\tilde{r}_{diss}\simeq10$.  From Fig \ref{fig-rpi} we deduce that at $\tilde{r}_{diss}$ 
the photopion depth $\tau_{p\gamma}^{syn}$ exceeds unity
at energies above $10^{15}$ eV, where $\xi_{syn}/A=0.1$ has been adopted.  
For a maximum proton energy of $10^{16}$ eV we then obtain $\bar{f}_\pi\simeq0.2$, 
and $N_{\mu}\simeq 0.01(\xi_{p}/0.1)$ for a 1 km$^2$ 
detector.  At $\gtrsim$10 TeV this is comparable to the background, but would require
an average of $\sim100$ outbursts for detection.  However, if a similar 
event were to occur in a source that is located closer to Earth, at a 
distance of say 3 kpc as in the case of the superluminal source 
GRO J1655-40 (which has a similar Doppler factor), or from a jet oriented at 
a smaller angle, then a few or even a single outburst may produce several 
muon events in such a detector.  Estimates of VHE neutrino fluxes for 
identified microquasars are listed in Ref.~\refcite{Distefano02}.

\subsection{GRBs}
\label{sec:application-GRB}
As already mentioned in \S~\ref{sec:obs-GRB}, the standard view (see Ref.~\refcite{Piran05} and references therein), at least until recently, was that the prompt GRB emission is produced behind internal shocks that form in the coasting region of a baryon loaded fireball.
This scenario was originally motivated by the detection, in many sources, of a power law component
of the prompt emission spectrum extending up to energies well above the MeV peak,
in conflict with the thermal spectrum expected from an adiabatically expanding, baryon free 
fireball\cite{Paczynsky86}.   The radius above which internal shocks can form is roughly 
$\tilde{r}_{diss}\simgt10^5(\Gamma_{\infty}/300)^2$, where $\Gamma_\infty$ is the Lorentz factor at the coasting region
(see \S \ref{sec:jet-diss}), and the Thomson depth is roughly $\tau_T\simeq n_b^\prime\sigma_T r/\Gamma_\infty
\simeq10^{6.5}{\cal L}_{j12}\Gamma_{\infty,300}^{-3}\theta_{-1}^{-2}\tilde{r}^{-1}$, where Eq. (\ref{n_b}) has been
used.  Thus, $\tau_T\sim1$ at $r\sim$ a few times $r_{diss}$.  Consequently, the prompt emission in this model originates
from radii $r\simgt10^{12}$ cm.  

The difficulty with purely leptonic fireball models mentioned above may be alleviated if dissipation 
occurs during the outflow acceleration.
The claim that internal shocks must form in the coasting regime,
since different shells cannot catch up at radii where the fireball
is accelerating, applies only to conical geometries.
Obliqueness effects, inhomogeneous injection of pairs, and collision of the pair blobs with
the surrounding matter may all lead to formation of shocks at smaller radii, where 
the optical depth exceeds unity.   Bulk Comptonization may in this case
lead to a nonthermal extension of the spectrum well above the thermal peak.
It is quite likely that these shocks pass through a photosphere\cite{Ghisellini99,Guetta01},
in which case the nonthermal photons may escape the system before being thermalized. 

The post-SWIFT discoveries of a shallow afterglow phase at early times,
and fastly rising, large amplitude x-ray flares in the early afterglow phase
\cite{Burrows05} introduces new puzzles.  According to some interpretations (e.g., Ref.~\refcite{Fan05})
these observations imply prolonged activity of the central engine.  If true, then this may indicate that 
$\gamma$-rays are emitted during the prompt phase with very high efficiency, in the
sense that the remaining bulk energy is a small fraction of the total 
blast wave energy inferred from the afterglow emission.  Such episodes
can be most naturally explained as resulting from dissipation in a pure electron-positron plasma.
However, the possibility that the shallow phase results from variable micro-physical parameters and
does not require extremely high efficiency\cite{Granot06} cannot be ruled out.

If indeed a considerable fraction of the bulk energy is carried by baryons in the dissipation region, as
envisioned in the ``standard'' model, then UHECRs and VHE neutrinos may be produced with a fluence comparable
to the $\gamma$-ray fluence.
For a single GRB explosion Eq. (\ref{eq:dotNmu}) yields an integrated number of
\begin{equation}
N_{\mu}\simeq 2\xi_{p}\bar{f}_\pi{\left({\Gamma\over100}\right)^2}{E_{j51}
\over d_{L28}^2}(A/1{\rm km}^2),
\label{eq:Nmu-GRB}
\end{equation}
muon events, where $E_j=10^{51}E_{j51}$ ergs is the total energy release in the explosion.  The neutrinos
should arrive within a time window of tens of seconds from the beginning of the explosion.
Typically $\bar{f}\sim0.1$ for long GRBs that exhibit extension of the prompt spectrum well
above the MeV peak, and may be larger for bursts which are optically thick to pair production
at these energies\cite{Waxman/Bahcall97}.  With $\xi_p\sim\bar{f}_\pi\sim0.1$ and $\Gamma\simeq300$, 
Eq. (\ref{eq:Nmu-GRB}) gives about $0.1$ events for a typical long GRB at a redshift 
$z=1$.  Consequently, only nearby GRBs can be individually detected by the upcoming experiments.  The cumulative 
flux produced by the entire GRB population has been estimated in Ref.~\refcite{Waxman/Bahcall97}
under the assumption that cosmological GEBs are the sources of the observed UHECRs.  This requires that the
rate of energy production of UHECRs in the energy interval $10^{19}-10^{21}$ eV is comparable to 
the rate of $\gamma$-ray production by GRBs in the BATSE band, $\dot{E}\sim10^{44.5}$ 
erg Mpc$^{-3}$ yr$^{-1}$ \cite{Waxman95}.  The neutrino flux thereby computed is comparable to
the background.  However, owing to the correlation in time and direction of the GRB neutrinos 
with the associated $\gamma$-ray sources, the signal should be easily detected above the background,
at a rate of about $10-100$ correlated events per year.

\section*{Acknowledgment}
I thank C. Dermer and E. Nakar for useful comments.
This work was supported by an ISF grant for the Israeli Center for High Energy Astrophysics


\end{document}